\documentclass{article}

\usepackage{amsmath,amssymb}
\usepackage{subfigure}
\usepackage{float}
\usepackage{graphicx}
\usepackage{color}
\usepackage{booktabs}
\usepackage{multirow}

\begin{document}




\title{Flow-priority optimization of additively manufactured variable-TPMS lattice heat exchanger based on macroscopic analysis}


\author{Kazutaka Yanagihara
\thanks{Department of Applied Mechanics and Aerospace Engineering, Waseda University, 59-311, 3-4-1 Okubo, Shinjuku-ku, Tokyo 169-8555, Japan}
\thanks{AGC Inc., 1-1 Suehiro-cho, Tsurumi-ku, Yokohama-shi, Kanagawa 230-0045, Japan}
\and
Jun Iwasaki$^*$
\and
Kiyoto Saso$^*$
\and
Taichi Yamashita$^*$
\and
Shomu Murakoshi$^*$
\and
Akihiro Takezawa$^*$
\thanks{Corresponding author, atakezawa@waseda.jp}
}
\date{} 
\maketitle

\begin{abstract}
Heat exchangers incorporating triply periodic minimal surface (TPMS) lattice structures have attracted considerable research interest because they promote uniform flow distribution, disrupt boundary layers, and improve convective heat transfer performance. However, from the perspective of forming a macroscopic flow pattern optimized for heat exchange efficiency, a uniform lattice is not necessarily the optimal configuration. This study presented a macroscopic modeling approach for a two-fluid heat exchanger equipped with a TPMS Primitive lattice.
Macroscopic flow analysis was conducted based on the Darcy--Forchheimer theory.
Under the assumption that heat is transferred solely at the interface between the fluid and TPMS walls, a macroscopic heat transfer model was developed using a volumetric heat-transfer coefficient, which serves as an artificial property characterizing the unit-volume heat transfer capability. To effectively regulate the relative dominance of the hot and cold flows and the channel widths within the heat exchanger, we adopted the isosurface threshold of the TPMS Primitive lattice as the design variable and constructed an optimization scheme for the lattice distribution using a previously described macroscopic model. Optimization was subsequently performed for a planar heat exchanger, where the hot and cold fluids followed U-shaped flow trajectories. The optimal solution was verified, and its validity was examined through detailed geometric analysis and experiments conducted using metal-based laser powder bed fusion. The optimal solution derived from the macroscopic model demonstrated a clear performance improvement over a uniform lattice, with an average enhancement of 24.2\% in the detailed-geometry simulations and 23.3\% in the experimental results.
\end{abstract}




\section{Introduction}
Recent advances in metal additive manufacturing (AM) have profoundly affected the design and evolution of mechanical parts \cite{frazier2014,herzog2016}. Owing to its capability to fabricate intricate internal geometries, metal AM enables the embedding of efficient coolant channels within a part, thereby enabling high-performance heat exchange structures. In practical settings, AM has been increasingly adopted to produce tooling inserts and cooling systems that feature complex channel networks, which is often referred to as conformal cooling \cite{shinde2017,feng2021}. Similarly, metal AM has been applied to the fabrication of two-fluid heat exchangers \cite{kaur2021,niknam2021}.

A distinctive feature of AM is its ability to create lattice architectures that incorporate hollow internal regions. Because lattice structures possess extremely large surface areas, they offer the potential for highly efficient cooling, and early studies frequently reported their application in conformal cooling \cite{wong2007, wong2009, brooks2016, kanbur2020, tan2020}. Among the wide variety of lattice configurations, triply periodic minimal surface (TPMS) lattices have become particularly prominent because they are mathematically described as surfaces that repeat periodically in three dimensions while maintaining a uniform zero mean curvature \cite{schoen1970}. The inherent continuity and periodicity of TPMS structures generate a uniform flow distribution, sustained boundary-layer disruption, and secondary flows, enabling improved convective heat transfer performance without the excessive pressure losses typically associated with turbulent-enhancing geometries. Therefore, studies on two-fluid heat exchangers that use TPMS lattices have been actively conducted in recent years \cite{kim2020,li2020,alteneiji2022,iyer2022,li2022,gao2023,liang2023,liang2023a,mahmoud2023,oh2023,reynolds2023,rover2023,wang2023,yan2023,song2024,wang2024,yan2024,oh2025,qian2025,xiao2025,yan2025ijhmt,yan2025ijts,zhang2025,zhang202502,zhao2025,zou2025}.

This study focuses on the flow channels of heat exchangers. First, studies have been conducted on heat exchangers in which cells are arranged linearly and heat exchange is achieved through straight counterflow streams \cite{kim2020, iyer2022, li2022, wang2023, yan2023, song2024, zhang2025}. Heat exchangers in which straight flow channels intersect at right angles have also been extensively studied \cite{alteneiji2022, gao2023, liang2023, liang2023a, reynolds2023, rover2023, yan2024, qian2025, xiao2025, yan2025ijhmt, zhang202502, zhao2025}. In these cases, the role of the TPMS lattice is to induce secondary flows and enhance convective heat transfer, functioning essentially as a flow filter. The flow within a TPMS lattice is macroscopically straight; thus, the lattice does not need to provide any function to guide or redirect the flow. In such situations, while the optimization of various parameters and operating conditions may still be important, actively optimizing the flow path geometry itself may not be necessary.

However, in some heat exchangers, one or both fluids follow U- or L-shaped flow paths \cite{li2020,mahmoud2023,oh2023,wang2024,oh2025,yan2025ijts,zou2025}. In such complex flow paths, we must consider not only the generation of local turbulence by the TPMS lattice geometry but also its influence on the macroscopic flow behavior. It is likely that, from the perspective of forming a macroscopic flow pattern optimized for heat-exchange efficiency, a uniform lattice is not necessarily the optimal configuration.

Li et al. concluded that optimizing the lattice wall thickness and porosity is essential for further improving the performance of TPMS lattice heat exchangers \cite{li2020}. Similarly, Wang et al. investigated a heat exchanger incorporating U-shaped flow paths and found that a uniformly distributed lattice resulted in different channel structures depending on the relative positions of the inlet and outlet, leading to asymmetric flow resistance; based on this observation, they emphasized the necessity of flow-path optimization \cite{wang2024}. Oh et al. introduced a mathematical filtering strategy that locally alters the TPMS lattice geometries to designate the inlet and outlet regions, embed flow-directional barriers, and reduce the solid fractions near the openings to lower the flow resistance. Using this approach, they achieved a heat exchange performance that was nearly double that of conventional heat exchangers \cite{oh2023}. However, in their work, the placement of these filters was determined heuristically, indicating that further enhancements require a fundamental numerical optimization method for the TPMS lattice geometry.

A major benefit of lattice structures is their ability to emulate the functionally graded material behavior by altering the lattice geometry according to the spatial location \cite{zhang2015, clausen2016}. Therefore, controlling the flow by introducing nonuniform geometries or density distributions is not difficult. Oh et al. introduced a design strategy for a gyroid-based heat exchanger in which the cell size was increased in low-velocity regions and decreased in high-velocity regions to achieve a more uniform flow distribution, thereby demonstrating an improvement in heat exchange efficiency through a graded lattice configuration \cite{oh2025}. However, this design was developed empirically, and the rigorous optimization of TPMS graded-lattice heat exchangers based on numerical computation remains an open problem. Furthermore, in TPMS lattices, the wall position can be easily modified by adjusting the parameters of the implicit surface representation, which allows the local flow priority between the hot and cold fluids to be readily controlled. However, to the best of the authors' knowledge, no studies have optimized the relative priority of the two flows in heat exchangers using TPMS lattices.

The design of lattice structures with spatial variation is accomplished more effectively using an approximate structural optimization method that utilizes homogenized material properties and applies gradient-driven optimization. Employing numerically homogenized effective properties \cite{guedes1990,andreassen2014} avoids the substantial computational cost that would otherwise arise from explicitly resolving the detailed lattice geometry. In this framework, the design variable is defined as either the effective density or a characteristic geometric parameter of the lattice, and it is iteratively updated using a gradient-based algorithm once the sensitivities have been computed. Consequently, the procedure closely resembles that used in conventional topology optimization \cite{bendsoe1988,bendsoe2003}. Topology optimization methods for two-fluid heat exchangers have also been proposed \cite{hoghoj2020,kobayashi2021}. However, topology optimization primarily aims at clearly defining the optimal distribution of fluid and solid regions, and the geometric complexity of the resulting structure is directly linked to computational cost. Consequently, it is difficult to generate intricate detailed geometries such as TPMS lattices within a topology optimization framework. Instead, such methods are generally intended to determine the main flow pathways in channel-type configurations. Therefore, although the distribution optimization of lattice structures with spatial variation and topology optimization share algorithmic similarities, the target structures they seek to achieve are fundamentally different.

The authors proposed an optimization framework that adjusts the lattice density within a heat exchanger to realize efficient cooling under liquid flow conditions characterized by Reynolds numbers of the order of $10^{2}$ \cite{takezawa2019am, takezawa2019ijhmt, takezawa2024, yanagihara2025}. In these analyses, fluid behavior was modeled using the Darcy--Forchheimer theory \cite{ward1964, joseph1982, nield2013}. Based on the macroscopic porous-flow approximation method, we propose an approach for optimizing the flow priority between two channels in a TPMS lattice heat exchanger. For the TPMS lattice, we employed the Primitive lattice, prioritizing its orthogonally symmetric geometry and isotropic effective properties, which simplified the modeling process. Under the assumption that heat is transferred solely at the interface between the fluid and TPMS walls, a macroscopic heat transfer model was developed using a volumetric heat-transfer coefficient, which serves as an effective parameter characterizing the unit-volume heat transfer capability \cite{younis1993}.
Using this volumetric heat-transfer coefficient together with the effective thermal conductivity and Darcy velocity, we derived the advection-diffusion heat transfer equations for each fluid. By simultaneously solving a set of five governing equations, including the heat conduction equation for the TPMS walls, we obtained the macroscopic velocity and temperature fields for each fluid along with the temperature distribution inside the solid walls.

Based on the macroscopic analysis described earlier, we treated the parameter that determined the wall position of the TPMS Primitive lattice as the design variable, which allowed us to optimize the relative flow importance of the hot and cold channels in each cell. The effective properties required in the macroscopic model were computed using representative volume element (RVE) based homogenization. In this study, optimization was applied to a counter-flow heat exchanger, where each fluid on the hot and cold sides moved through a U-shaped channel. Such U-shaped counterflow arrangements present notable design challenges and have been the subject of numerous prior investigations on flow-path topology optimization \cite{hoghoj2020,kobayashi2021}.

In Section 2, we describe the analysis method, optimization procedure, and experimental approach used in this study. Section 3 presents the optimization results, followed by a reanalysis using detailed simulations and corresponding experimental validations. In Section 4, we examine the soundness of the proposed analytical approach, explore the mechanical principles underlying the optimal solutions, and evaluate the extent to which the optimized design can be reproduced experimentally.

\section{Methods}
\subsection{Macroscopic analysis model}
\subsubsection{Governing equations}
We employed a TPMS lattice structure as the internal geometry of the two-channel heat exchanger. By optimally distributing the isosurface threshold of the TPMS, we aim to control the flow priority of each fluid, thereby maximizing the overall heat-exchange performance. 

At the inlet boundary $\Gamma_{\mathrm{in}}^{fh}$, the incompressible hot fluid is supplied at pressure $P_{\mathrm{in}}^{fh}$ and temperature $T_{\mathrm{in}}^{fh}$, while at the outlet boundary $\Gamma_{\mathrm{out}}^{fh}$, it discharges under the condition $P^{fh} = 0\,\mathrm{Pa}$.
Similarly, incompressible cold fluid enters from $\Gamma_{\mathrm{in}}^{fc}$ with pressure $P_{\mathrm{in}}^{fc}$ and temperature $T_{\mathrm{in}}^{fc}$, and leaves through $\Gamma_{\mathrm{out}}^{fc}$ with $P^{fc} = 0\,\mathrm{Pa}$. In this study, these inlet temperatures are imposed as thermal boundary conditions, while the outlet temperatures are obtained as part of the solution. The two fluids exchange heat through the solid walls of the embedded TPMS lattice within the heat exchanger and are assumed not to mix with each other. The outer walls of the heat exchanger were treated as adiabatic. After solving the local flow and temperature fields, the overall heat-exchange rate of the heat exchanger was evaluated from the enthalpy difference between the inlet and outlet boundaries of each fluid. For simplicity, both fluids were assumed to be composed of the same material.

\begin{figure}[H]
\centering
\includegraphics[scale=0.8,clip]{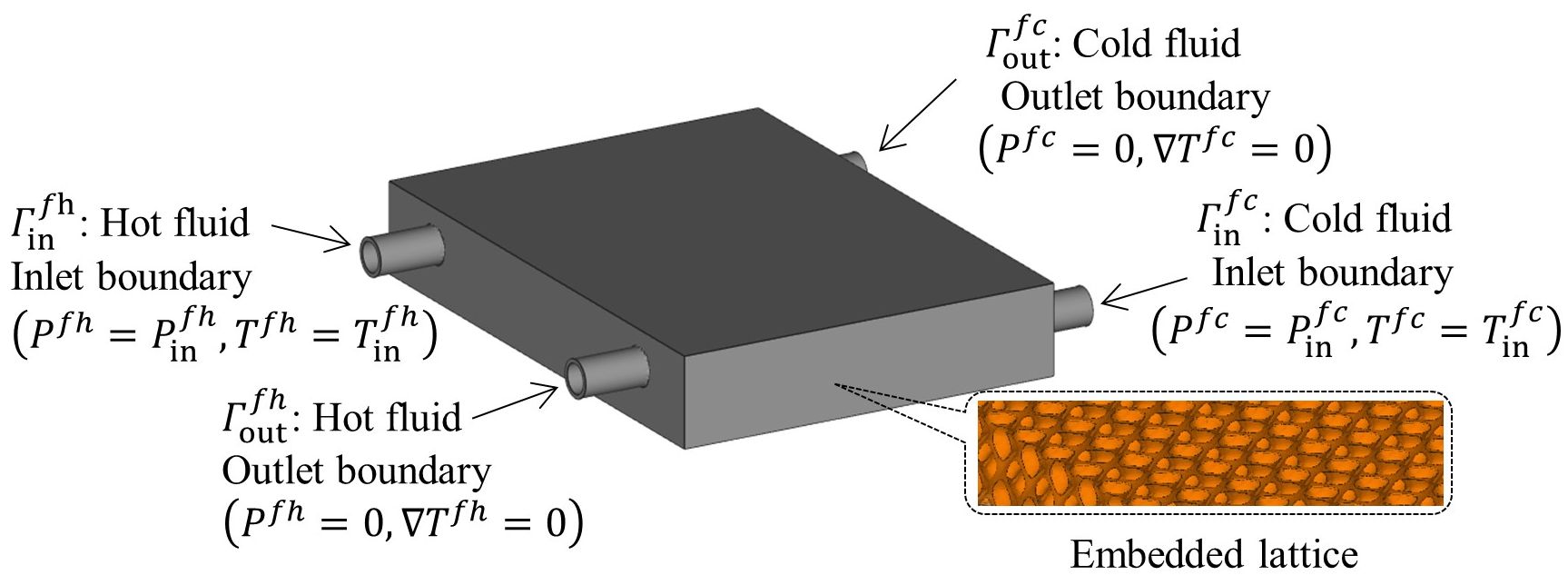}
\caption{Outline of analysis model.}
\label{outline}
\end{figure}

However, during the iterative optimization process, repeatedly reconstructing the detailed lattice geometry, in which the two fluid domains and the separating solid walls are explicitly resolved, and performing full thermo-fluid simulations would be computationally prohibitive. Therefore, to substantially reduce the computational cost, the lattice structure was treated macroscopically as a porous medium, and an approximate porous-flow model based on effective material properties was employed. For clarity, we distinguish between two levels of modeling in this study. We first present the governing equations of the detailed model, in which the TPMS geometry is explicitly resolved. We then introduce the macroscopic porous-medium model used in the optimization.

The detailed model is governed by the Navier--Stokes and continuity equations for each fluid, together with the energy equation for heat transfer in the fluid and solid domains.
\begin{gather}
\rho \left( \mathbf{u}^{(\alpha)} \cdot \nabla \mathbf{u}^{(\alpha)} \right)
= - \nabla P^{(\alpha)} + \mu \nabla^{2} \mathbf{u}^{(\alpha)}
\label{eq01}\\
\nabla \cdot \mathbf{u}^{(\alpha)} = 0 \label{eq02}\\
\rho C_{p} \, \mathbf{u}^{(\alpha)} \cdot \nabla T^{(\alpha)}
- \lambda^{(\alpha)} \nabla^{2} T^{(\alpha)} = Q^{(\alpha)}, \label{eq03}\\
- \lambda^{s} \nabla^{2} T^{s} = Q^{s}, \label{eq03-2}\\
\qquad (\alpha = fh, fc) \nonumber
\end{gather}
where $\mathbf{u}$ and $P$ represent the velocity and pressure of each fluid, respectively; $\rho$ is the fluid density; $\mu$ is the dynamic viscosity; $T$ is the temperature; $C_{p}$ is the specific heat at constant pressure; $\lambda$ is the thermal conductivity; and $Q$ corresponds to the volumetric internal heat source; the superscripts $fh$, $fc$, and $s$ denote the hot-fluid domain, cold-fluid domain, and solid domain, respectively.

Thus, Eqs. \eqref{eq01}-\eqref{eq03-2} describe the detailed-geometry model, in which the two fluid domains and the intervening solid wall of the TPMS lattice are explicitly resolved. Because directly solving this model throughout the optimization would be computationally prohibitive, the optimization is instead performed using the following macroscopic porous-medium model.

When the lattice structure is treated macroscopically as a porous medium, the relationship between the pressure gradient and flow velocity can be expressed by the Darcy--Forchheimer law as given below \cite{ward1964, joseph1982, nield2013}.
\begin{equation}
\nabla \overline{P} = - \frac{\mu}{\kappa} \, \overline{\mathbf{u}} - \beta \rho \, |\overline{\mathbf{u}}| \, \overline{\mathbf{u}}
\label{eq04}
\end{equation}
where $\kappa$ and $\beta$ denote permeability and drag coefficients, respectively. The quantity $\overline{\mathbf{u}}$ denotes the Darcy velocity and $\overline{P}$ indicates the macroscopic pressure.
By combining Eqs. \eqref{eq01} and \eqref{eq04} in an analogous manner, the flow in the macroscopic model can be expressed using the Brinkman--Forchheimer equation with porosity $\varepsilon$ of the porous medium \cite{nield2013, vafai1981, vafai1982, hsu1990}:

\begin{equation}
\frac{\rho}{\varepsilon}\, \overline{\mathbf{u}} \cdot 
\left( \nabla \frac{\overline{\mathbf{u}}}{\varepsilon} \right)
= 
- \nabla \overline{P}
+ \frac{\mu}{\varepsilon} \nabla^{2} \overline{\mathbf{u}}
- \frac{\mu}{\kappa} \overline{\mathbf{u}}
- \beta \rho \, |\overline{\mathbf{u}}| \, \overline{\mathbf{u}}
\label{eq05}
\end{equation}

In a TPMS lattice, two fluid channels are separated by solid walls. Consequently, within the macroscopic framework, the Brinkman--Forchheimer and continuity equations were written independently for each fluid. Thus, for hot fluid $fh$ and cold fluid $fc$, these governing equations are expressed as follows:

\begin{gather}
\frac{\rho}{\varepsilon^{(\alpha)}} \, \overline{\mathbf{u}}^{(\alpha)} \cdot 
\left( \nabla \frac{\overline{\mathbf{u}}^{(\alpha)}}{\varepsilon^{(\alpha)}} \right)
= 
- \nabla \overline{P}^{(\alpha)}
+ \frac{\mu}{\varepsilon^{(\alpha)}} \nabla^{2} \overline{\mathbf{u}}^{(\alpha)}
- \frac{\mu}{\kappa^{(\alpha)}} \overline{\mathbf{u}}^{(\alpha)}
- \beta^{(\alpha)} \rho \, \left| \overline{\mathbf{u}}^{(\alpha)} \right| \, \overline{\mathbf{u}}^{(\alpha)}
\label{eq06}\\
\nabla \cdot \overline{\mathbf{u}}^{(\alpha)} = 0
\label{eq07}\\
\qquad (\alpha = fh, fc)
\nonumber
\end{gather}

However, for the thermal field, the heat advection-diffusion equation in the macroscopic model is obtained by replacing the velocity in Eq. \eqref{eq03} with the Darcy velocity and by substituting the effective thermal conductivity for the original thermal conductivity. 
In this formulation, the temperatures of the fluids $fh$ and $fc$, and the solid are treated as distinct variables within a single macroscopic lattice. 
Accordingly, the macroscopic temperature distributions follow the three equations given below:

\begin{gather}
\rho C_{p}\, \overline{\mathbf{u}}^{(\alpha)} \cdot \nabla \overline{T}^{(\alpha)}
= \overline{\lambda}^{(\alpha)} \nabla^{2} \overline{T}^{(\alpha)} - \overline{Q}^{(\alpha)\rightarrow s}
\label{eq08}\\
\overline{\lambda}^{\,s} \nabla^{2} \overline{T}^{\,s}
+ \sum_{i} \overline{Q}^{(\alpha)\rightarrow s} = 0
\label{eq09}\\
\qquad (\alpha = fh, fc)
\nonumber
\end{gather}

where $\overline{\lambda}$ and $\overline{T}$ denote the effective thermal conductivity and the macroscopic temperature of each phase in the porous-medium model, respectively; $\overline{T}$ is therefore an averaged quantity by definition; and $\overline{Q}^{(\alpha)\rightarrow s}$ represents the volumetric heat transfer from fluid $\alpha$ to the solid phase.

As described below, the macroscopic description of the heat transfer employs a volumetric heat-transfer coefficient $\overline{h}$, which represents the rate of thermal exchange between the solid and each fluid \cite{younis1993}.

\begin{gather}
\overline{Q}^{(\alpha)\rightarrow s}
= \overline{h}^{(\alpha)\rightarrow s} \left(\overline{T}^{(\alpha)}-\overline{T}^s \right)
\label{eq10}\\
\qquad (\alpha = fh, fc)
\nonumber
\end{gather}

\subsubsection{Derivation of effective parameters}

\begin{figure}[H]
\centering
\includegraphics[scale=0.8,clip]{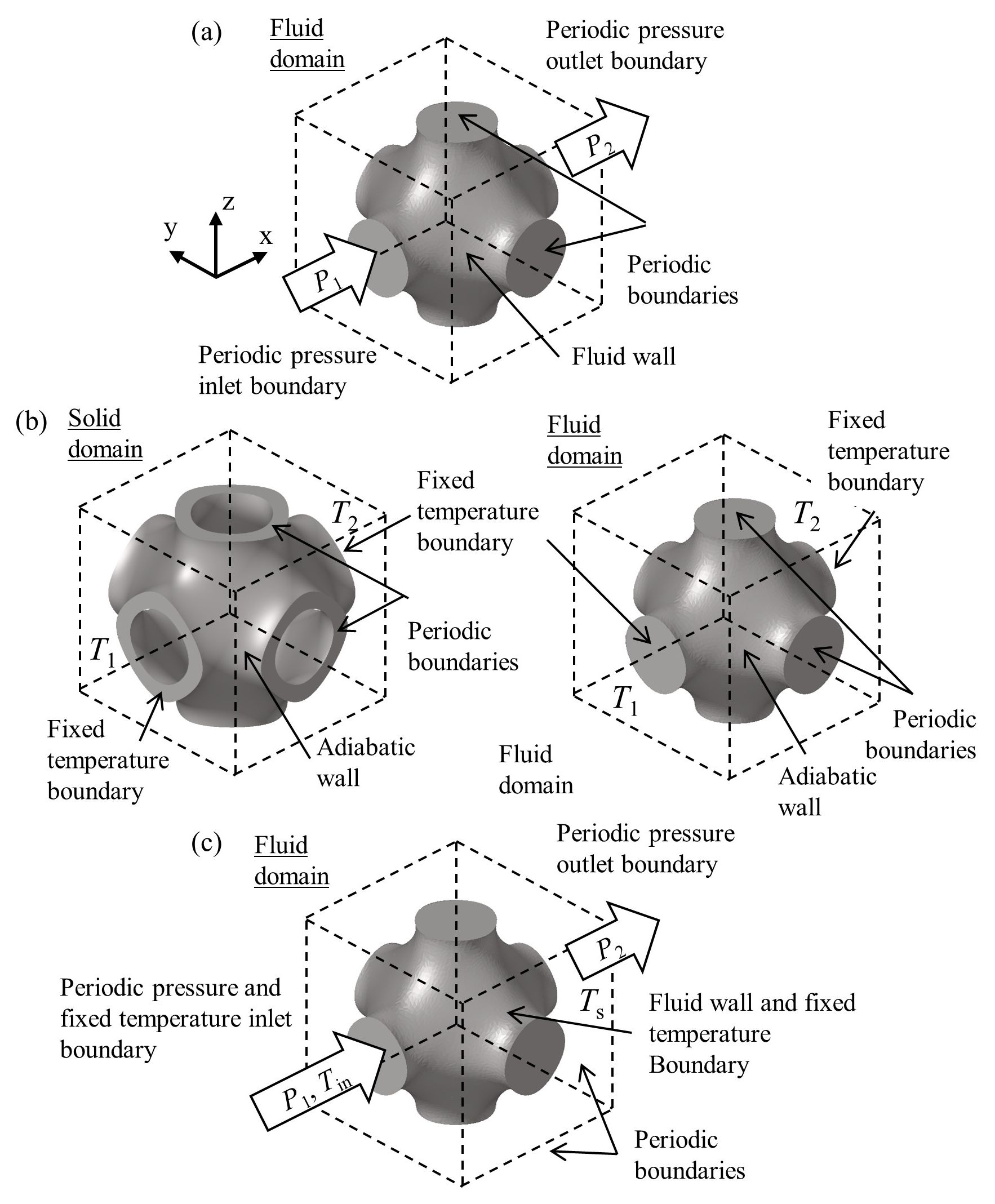}
\caption{Analysis models and boundary conditions for calculating the effective physical properties: (a) permeability and drag coefficient, (b) effective heat conductivity, and (c) volumetric heat-transfer coefficient.}
\label{rve}
\end{figure}

The effective parameters used in the macroscopic model were derived by RVE-based homogenization, where the flow and thermal states in the unit cell were obtained by solving the microscopic governing equations, i.e., Eqs.~\eqref{eq01}--\eqref{eq03-2}. The permeability and drag coefficients in Eqs. \eqref{eq06} and \eqref{eq07} are obtained by solving the Navier--Stokes equations in Eqs. \eqref{eq01} and \eqref{eq02} under boundary conditions that impose a pressure gradient, using a unit-cell structure corresponding to one period of the TPMS lattice, as illustrated in Fig. \ref{rve} (a). Although the pressure distribution and velocity field inside the unit cell were not uniform, they were treated as macroscopically uniform. The average outlet velocity is regarded as the Darcy velocity and is substituted into Eq. \eqref{eq04} to determine the permeability and drag coefficient. 
The porosity was calculated geometrically as the ratio of the fluid volume to the total unit-cell volume of the lattice structure.

Similarly, effective thermal properties were derived using the RVE method. The effective properties required in Eqs.~\eqref{eq08}--\eqref{eq10} are the effective thermal conductivities $\overline{\lambda}^{(\alpha)}$ $(\alpha = fh, fc, s)$ and the volumetric heat-transfer coefficients $\overline{h}^{(\alpha)\rightarrow s}$ $(\alpha = fh, fc)$. The effective thermal conductivity $\overline{\lambda}^{(\alpha)}$ was obtained from the unit-cell model shown in Fig.~\ref{rve}(b) under no-flow conditions by solving a steady pure-conduction problem. In this calculation, neither the Navier--Stokes equations nor the advective term was considered. A macroscopic temperature gradient was imposed in a prescribed direction, and the resulting average heat flux $\overline{\boldsymbol{J}}$ was evaluated from the total conductive heat flow crossing the pair of external unit-cell boundary faces perpendicular to that direction, divided by the area of those faces. Thus, $\overline{\boldsymbol{J}}$ is defined on the basis of the external unit-cell boundary faces, not the TPMS internal surface area. The effective thermal conductivity was then determined using the Fourier-type expression shown below:

\begin{equation}
-\overline{\lambda} \nabla T = \overline{\bold{J}}
\label{eq11}
\end{equation}

The permeability and drag coefficients, as well as the effective thermal conductivity, are macroscopic physical properties that depend only on the material properties and the microgeometry. In contrast, the volumetric heat transfer coefficient is an effective parameter that depends not only on the material properties and the microgeometry but also on the surrounding flow conditions. The volumetric heat-transfer coefficient $\bar{h}^{f \rightarrow s}$ was evaluated for the flow domain, as shown in Fig. \ref{rve} (c). 
As illustrated in Fig. \ref{rve} (c), the inlet temperature $T_{\mathrm{in}}^{f}$ is prescribed at the fluid inlet boundary, 
and the solid temperature $T^s$ was prescribed at the solid wall. 
Here, the solid temperature $T^s$ is assumed to be spatially uniform within the unit cell for the derivation of the volumetric heat-transfer coefficient. This approximation is introduced under the assumption that the characteristic dimension of the lattice unit cell is sufficiently small compared with the macroscopic length scale of the heat exchanger.

By solving the Navier--Stokes and heat advection--diffusion equations in the unit-cell domain, the fluid temperature field $T^f$ is obtained. The heat transferred from the fluid phase to the solid phase, $Q^{f \to s}$, is then evaluated from the inlet--outlet enthalpy balance of the fluid as follows:
\begin{equation}
Q^{f \to s}=\rho C_{p}
\left(
\int_{\Gamma_{\mathrm{in}}^{f}}
T_{\mathrm{in}}^{f}\mathbf{u}^{f} \cdot \mathbf{n}
\, dS
-
\int_{\Gamma_{\mathrm{out}}^{f}}
T^{f}\mathbf{u}^{f} \cdot \mathbf{n}
\, dS
\right)
\label{eq12}
\end{equation}
where $\Gamma_{\mathrm{in}}^f$ and $\Gamma_{\mathrm{out}}^f$ denote the inlet and outlet boundaries of the fluid domain, respectively, and $\mathbf{n}$ is the outward unit normal vector on each boundary.

As shown in Eq.~\eqref{eq10}, the volumetric heat-transfer coefficient is defined with respect to the macroscopic phase temperatures. In the present RVE-based derivation, the corresponding fluid temperature is evaluated as the explicit volume average of the microscopic temperature field $T^f$ obtained from the detailed unit-cell analysis, while the solid temperature is given by the prescribed uniform temperature $T^s$. Accordingly, $\bar{h}^{f \to s}$ is obtained by dividing the heat quantity by the temperature difference between the fluid and solid phases as follows:
\begin{equation}
\bar{h}^{f \to s} = \frac{Q^{f \to s}}{\displaystyle \left(\frac{1}{V^f}\int_{V^f} T^f dV-T^s\right)V}
\label{eq12.5}
\end{equation}
where $V^f$ and $V$ are the volume of the fluid domain and the unit cell, respectively. Thus, the temperature difference used in Eq.~\eqref{eq12.5} is consistent with that appearing in Eq.~\eqref{eq10}, although it is written here in its explicit unit-cell form. Because $T^f$ is linear with respect to $T_{\mathrm{in}}^{f}$ and $T^s$, $\bar{h}^{f \to s}$ depends on the velocity $\mathbf{u}^{f}$, which takes the same value on $\Gamma_{\mathrm{in}}^{f}$ and $\Gamma_{\mathrm{out}}^{f}$ owing to the periodic boundary conditions.

It should be noted that the effective parameters---the permeability, drag coefficient, and volumetric heat transfer coefficient---depend on the pressure difference used in the RVE analysis. Therefore, the pressure difference adopted in the RVE analysis should be determined in consideration of the pressure drop assumed in the optimization. In addition, if the pressure difference in each cell of the optimal design exceeds that assumed when deriving the effective parameters, the prediction accuracy may deteriorate.

\subsection{Optimization method}
\subsubsection{Formulation of the TPMS lattice and definition of design variables}

A TPMS is defined as a continuous function that partitions a domain into multiple disjointed regions. 
This position could be modified by changing the isosurface threshold value of the defining function. 
In this study, two TPMSs were defined, and the region between them was regarded as the solid lattice domain. 
The inner and outer regions were defined as the flow domains. 
We employed a TPMS represented using the primitive function as follows, prioritizing its orthogonally symmetric geometry and isotropic effective properties, which simplifies the modeling process:

\begin{equation}
\cos\left( \frac{2\pi}{l} x \right)
+ \cos\left( \frac{2\pi}{l} y \right)
+ \cos\left( \frac{2\pi}{l} z \right)
= C
\label{eq13}
\end{equation}

where $x, y, z$ are the spatial coordinates and $l$ is the length corresponding to one period of the Primitive function, which also corresponds to the unit cell size. 
In this study, $l$ was set to 5\,mm. 
The constant $C$ is the isosurface threshold that determines spatial partitioning. 
By using this Primitive function, the flow domains $\Omega^{fh}$, $\Omega^{fc}$, and the solid domain $\Omega^{s}$ are defined as follows:

\begin{gather}
\Omega^{fh}
=
\left\{
x, y, z
\mid
\cos\left( \frac{2\pi}{l} x \right)
+ \cos\left( \frac{2\pi}{l} y \right)
+ \cos\left( \frac{2\pi}{l} z \right)
< C - dC
\right\}\label{eq14}
\\
\Omega^{fc}
=
\left\{
x, y, z
\mid
\cos\left( \frac{2\pi}{l} x \right)
+ \cos\left( \frac{2\pi}{l} y \right)
+ \cos\left( \frac{2\pi}{l} z \right)
> C + dC
\right\}
\label{eq15}\\
\Omega^{s}
=
\left\{
x, y, z
\mid
\left|
\cos\left( \frac{2\pi}{l} x \right)
+ \cos\left( \frac{2\pi}{l} y \right)
+ \cos\left( \frac{2\pi}{l} z \right)
- C
\right|
\le dC
\right\}
\label{eq16}
\end{gather}

where $dC$ denotes a parameter that determines the thickness of the solid domain. Considering manufacturability, $dC = 0.3$ was chosen such that the minimum wall thickness did not fall below approximately 0.5\,mm.

\begin{figure}[H]
\centering
\includegraphics[scale=1.0,clip]{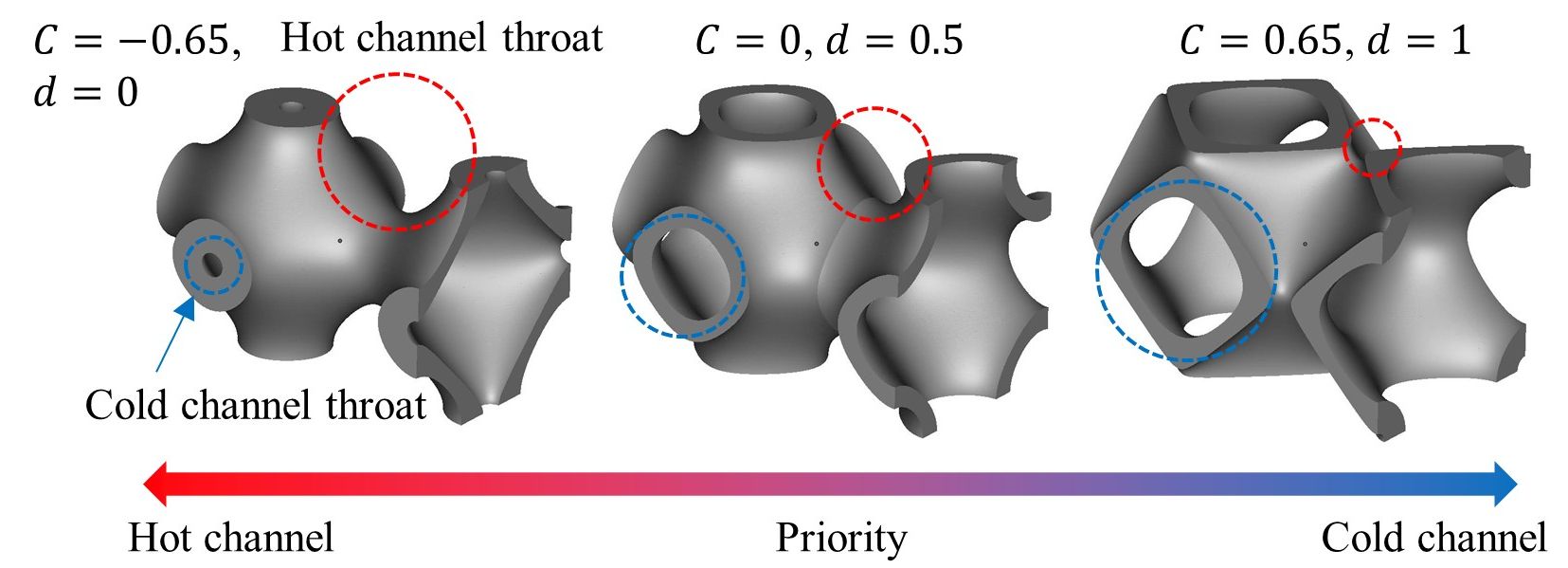}
\caption{Relationship between the value of the isosurface threshold $C$ and the lattice geometry}
\label{threshold}
\end{figure}

The isosurface threshold value $C$ of the TPMS function determines the position of the solid region as follows: 
By varying this value, the volumes of the flow regions and the thicknesses of the passages can be adjusted. 
In this study, we considered the problem of optimizing the flow priority using the isosurface threshold $C$ 
as the design variable. Fig. \ref{threshold} illustrates the change in the shape of the solid region when the threshold value was varied. 
In this study, we considered $C = 0$ as the reference value corresponding to equal flow resistances of the hot and cold channels. 
When $C > 0$, the high-temperature flow region dominates, whereas when $C < 0$, the low-temperature flow region dominates. 

The normalized design variable $d$ ($0 \le d \le 1$) is defined by normalizing the isosurface threshold $C$ as follows:
\begin{equation}
d = \frac{C - C_{\min}}{C_{\max} - C_{\min}}
\label{eq17}
\end{equation}
where $C_{\min}$ and $C_{\max}$ represent the lower and upper limits of the feasible range of $C$. 
To ensure that the minimum passage thickness does not fall below approximately 0.5 mm, we set $C_{\min} = -0.65$ and $C_{\max} = 0.65$. These bounds also function as implicit geometric constraints that prevent excessive narrowing of either the hot- or cold-fluid channel.

Although the design variable $d$ can be defined as the distance from the center of the unit cell to the isosurface, the TPMS value used to generate the lattice is a continuous scalar field distributed throughout space. The TPMS geometry is obtained by selecting a specific level-set value. Consequently, the actual TPMS lattice unit-cell geometry does not perfectly exhibit orthogonal symmetry, unlike the idealized shapes shown in Fig.~\ref{rve}. Therefore, the effective parameters derived using the RVE-based homogenization may include discrepancies when applied to the actual geometry. To address this issue, the validity of the macroscopic model is examined through detailed numerical analyses and experiments conducted using the fabricated lattice shapes.

\subsubsection{Optimization formulation}

\begin{figure}[H]
\centering
\includegraphics[scale=0.9,clip]{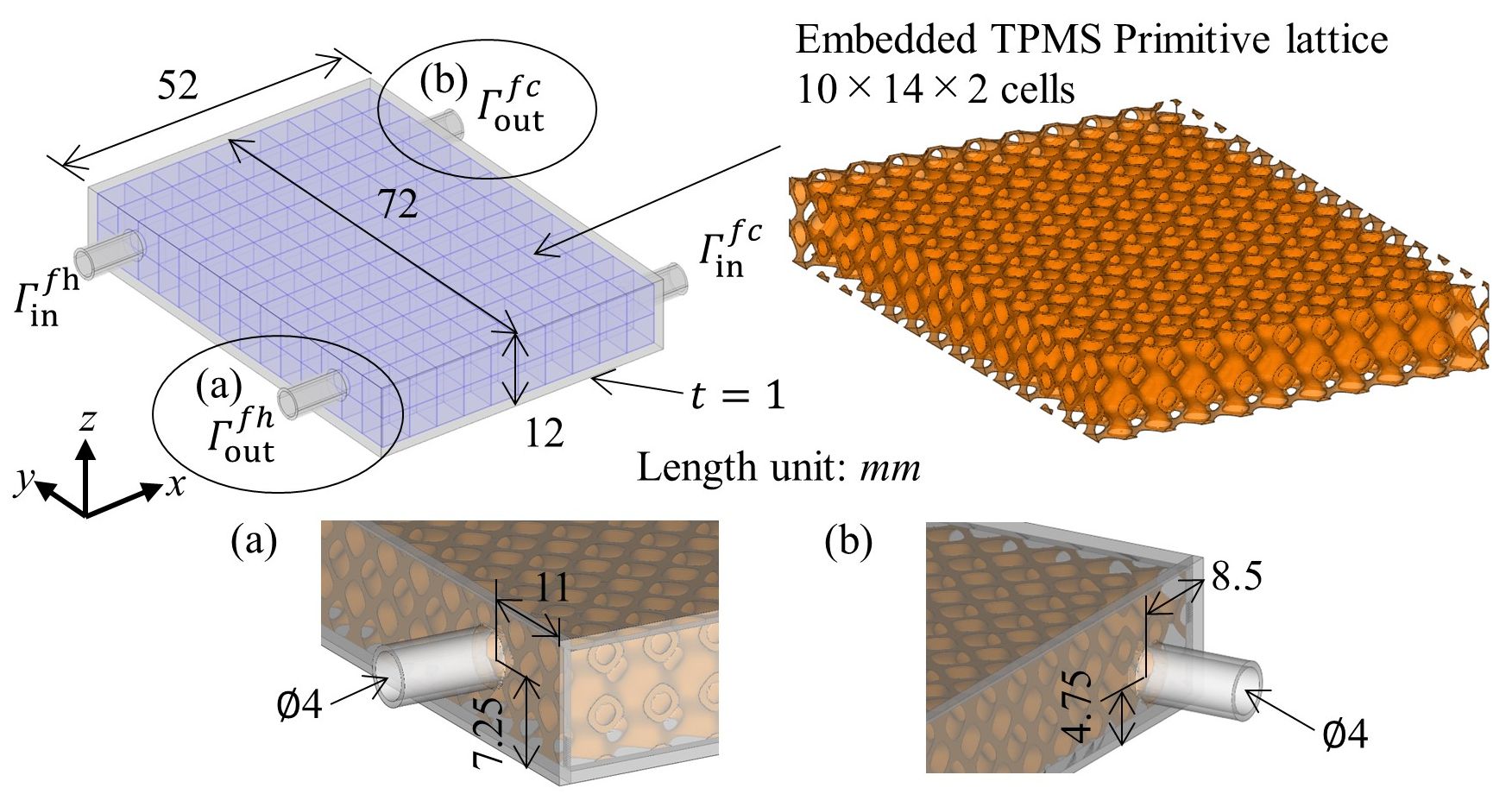}
\caption{Detailed analysis model for optimization.}
\label{detailedmodel}
\end{figure}

The details of the analysis model used for the optimization in this study are shown in Fig. \ref{detailedmodel}. 
All the exterior boundaries of the heat exchanger were modeled as adiabatic, meaning that heat transfer occurred only between the solid walls and flowing fluids. 
The working fluid on both the hot and cold sides was water with a mass density of 1000 kg/m$^{3}$, 
dynamic viscosity of 0.001 kg/(m$\cdot$s), specific heat of 4200 J/(kg$\cdot$K), 
and thermal conductivity of 0.6 W/(m$\cdot$K), respectively. 
The solid region is made of 316L stainless steel with a thermal conductivity of 14.1 W/(m$\cdot$K) \cite{simmons2020}.

In the macroscopic optimization model, the design variables are discretely defined at the centers of the cubic lattice cells with a spacing corresponding to one unit-cell period, and their values were linearly interpolated between these definition points. To prevent blockage of the flow passages at the inlet and outlet connections, which would lead to numerical instability in the simulations, the design variables of the lattice cells directly connected to the inlets and outlets were fixed. Specifically, the design variables at the cell centers connected to the hot-fluid inlet and outlet were set to $d=0$, whereas those connected to the cold-fluid inlet and outlet were set to $d=1$. Since such artificial fixing of the design variables may influence the optimal solution, the number of constrained cells was kept as small as possible. The total number of design variables was 280, of which 20 variables directly connected to the inlet and outlet regions were fixed. Therefore, the number of free design variables actually optimized was 260.

In this study, following our previous work, we assumed practical constraints such as the pump capacity in real applications, and imposed a pressure drop as a pressure boundary condition at the inlets and outlets to prevent the optimization process from producing solutions that result in excessive pressure losses \cite{takezawa2024,yanagihara2025}. 

The inlet temperature of the hot fluid was set to $60\,^\circ\mathrm{C}$, and that of the cold fluid was set to $25\,^\circ\mathrm{C}$. 
For both fluids, the pressure difference between the inlet and outlet was set to $500\,\mathrm{Pa}$. 
In addition, in the detailed geometric model, the reference coordinate position of the primitive function defining the TPMS was adjusted so that the openings of the lattice structure were aligned with the inlet and outlet ports.

The objective of this study was to maximize the heat transfer rate in the heat exchanger. 
Assuming that the heat exchanger is thermally insulated from the surroundings, the heat gained by the cold fluid and the heat lost by the hot fluid are equal. 
Therefore, the objective function of the optimization problem is defined as the maximization of the heat transfer rate, which is evaluated from the difference between the inlet and outlet enthalpy flow rates of the hot fluid, as expressed below:

\begin{equation}
\text{maximize}\ \rho C_{p}\left( \int_{\Gamma_{\mathrm{in}}^{fh}}  T^{fh} \mathbf{u}^{fh} \cdot \mathbf{n}dS - \int_{\Gamma_{\mathrm{out}}^{fh}}  T^{fh} \mathbf{u}^{fh} \cdot \mathbf{n}dS\right)
\label{eq18}
\end{equation}

where $S$ represents the boundary surface and $\mathbf{n}$ is the outward unit normal vector on the boundary. No additional constraints are imposed.

\subsubsection{Optimization algorithm}
\begin{figure}[H]
\centering
\includegraphics[scale=1.0,clip]{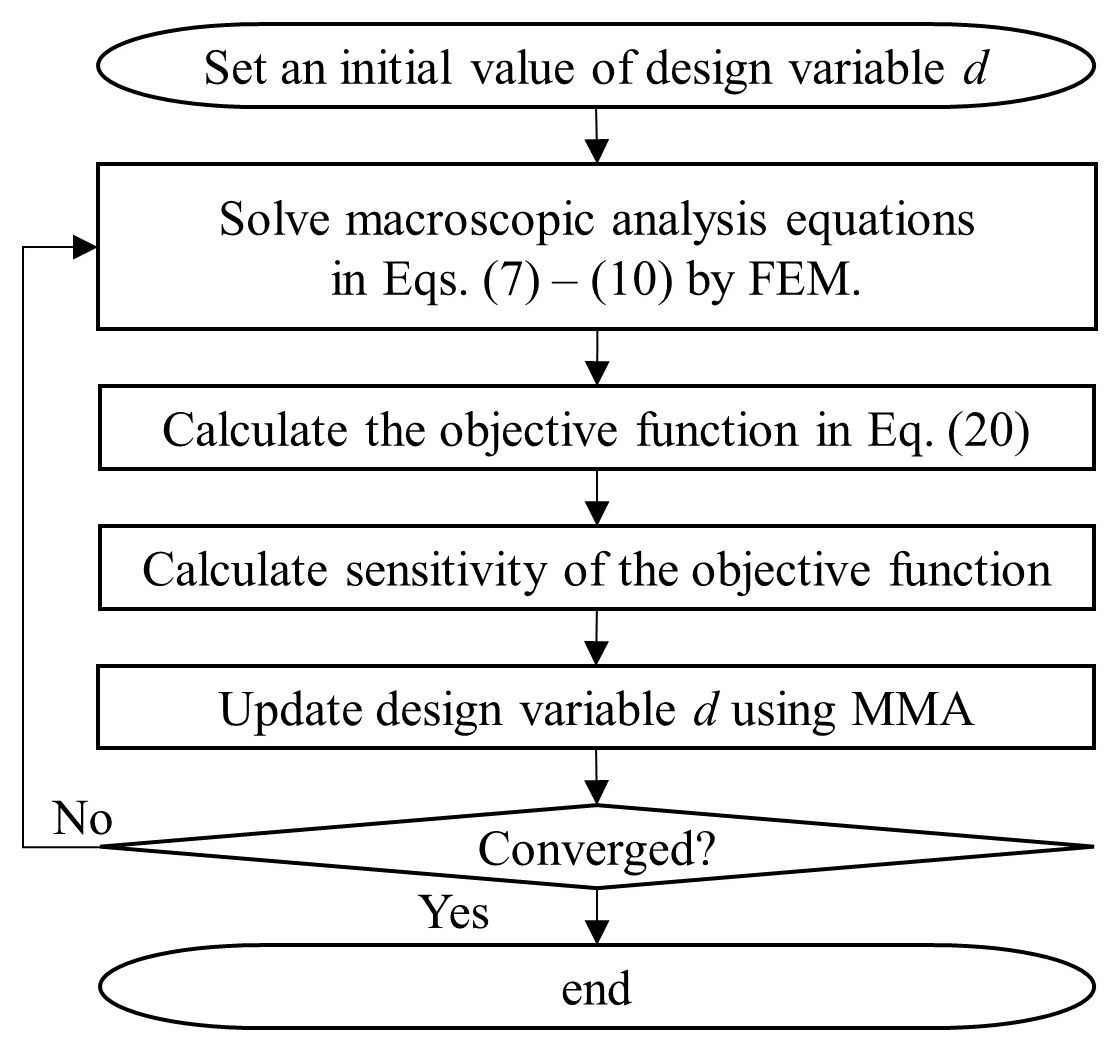}
\caption{Optimization algorithm flowchart.}
\label{flowchart}
\end{figure}

A flowchart of the optimization procedure is presented in Fig. \ref{flowchart}. 
Before starting the optimization, the design variable $d$ is initialized to $0.5$ over the entire design domain. 
Within each optimization loop, the Brinkman--Forchheimer equations in Eq. \eqref{eq06}, 
the continuity equation in Eq. \eqref{eq07}, the macroscopic heat advection-diffusion equations in Eq. \eqref{eq08}, 
and the macroscopic heat conduction equation in Eq. \eqref{eq09} are solved using finite element method (FEM). 
Subsequently, the objective function shown in Eq. \eqref{eq18} and its sensitivities are evaluated.
The design variables were updated using the optimization algorithm. 
The method of moving asymptotes (MMA) was employed as the update scheme \cite{svanberg1987}.

\subsubsection{Numerical simulation tools}
In this study, COMSOL Multiphysics, a finite element method solver, was used to derive the effective material properties and analyze the macroscopic model, while Star-CCM+, a finite volume method solver, was employed for reanalysis using detailed geometry.

\subsection{Experimental method}
\begin{figure}[H]
\centering
\includegraphics[scale=0.1,clip]{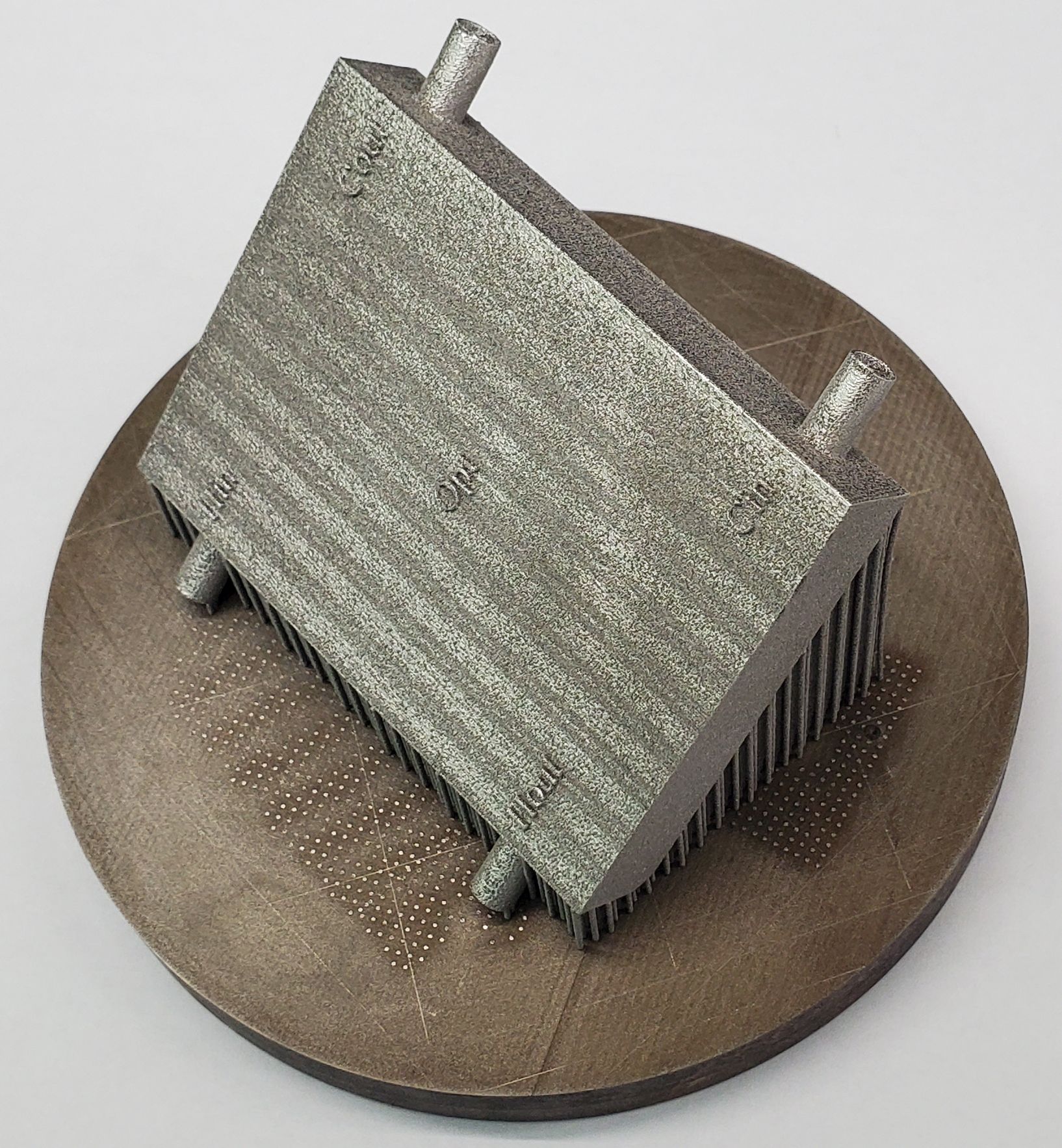}
\caption{Build orientation of the test piece.}
\label{testpiece}
\end{figure}

Heat exchanger specimens were fabricated using a metal laser powder bed fusion (LPBF) apparatus 
(ORLAS Creator RA, 2onelab Inc., Germany). The powder used was a 316 L stainless steel (H\"ogan\"as, Sweden) with a mean particle diameter of 30\,\textmu m. Nitrogen is used as the ambient gas. Table \ref{parameters} summarizes the process parameter settings of the device. The hatch spacing is set to be equal to the spot size. The laser power and scanning speed was adjusted to obtain a mass density of more than 99\% for the 10 mm square solid cube test pieces made of 316 L stainless steel. The layer thickness is set to be smaller than the average powder diameter because the ORLAS Creator RA uses a rubber blade for powder feeding. As shown in Fig. \ref{testpiece}, the specimen was fabricated at an angle of $45^\circ$ with respect to the horizontal plane, and support structures were applied only to the outer surfaces of the specimen. No heat treatment or polishing was performed after fabrication. A leakage test was conducted by flowing water through each channel separately in the as-fabricated specimen to confirm that no leakage occurred between the hot and cold channels.

\begin{table}[H]
\centering
\caption{Process parameters of metal LPBF. }
\label{parameters}
\footnotesize
\begin{tabular}{cc}
Parameter name&Value\\
\hline
Laser power [W]&132\\
Scan speed [mm/s]&400\\
Beam diameter [$\mu$m]&40\\
Hatching spacing [$\mu$m]&40\\
Layer thickness [$\mu$m]&25\\
\hline
\end{tabular}
\end{table}

\begin{figure}[H]
\centering
\includegraphics[scale=1.0,clip]{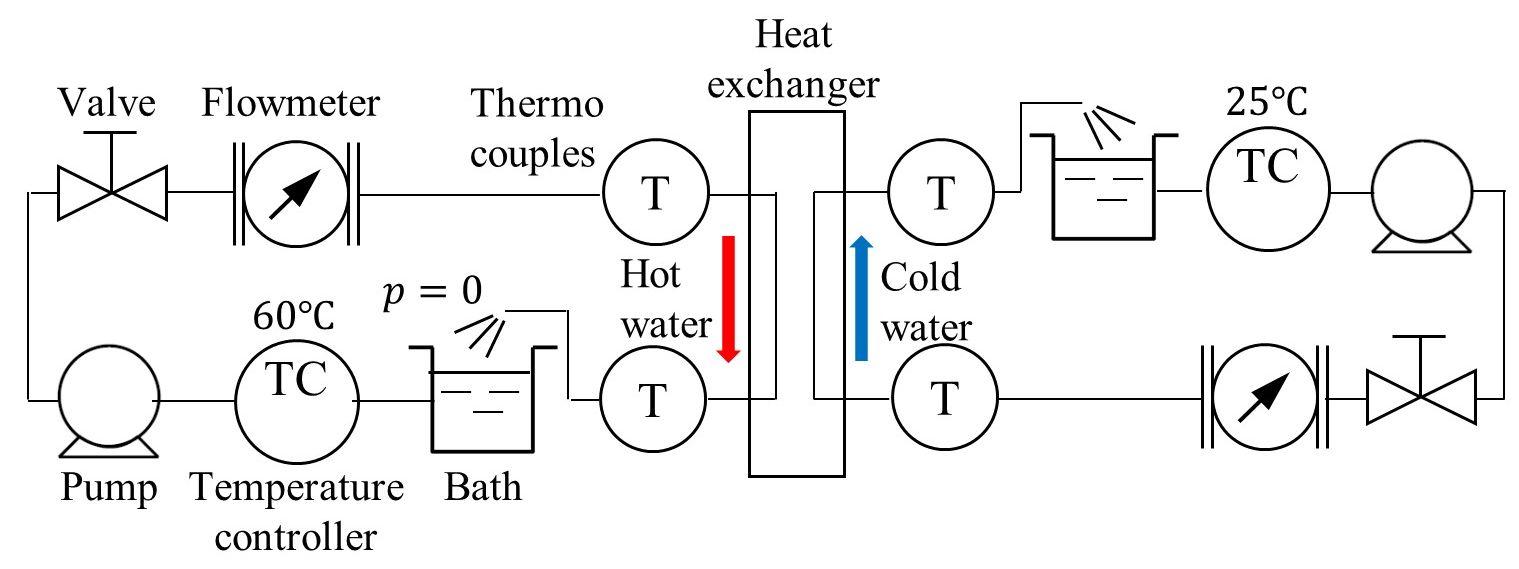}
\caption{Flow diagram of experimental setup.}
\label{expsetup}
\end{figure}

A schematic of the experimental setup is shown in Fig. \ref{expsetup}. 
Chilled water was delivered by using a cooling thermostat unit (LTC-1200$\alpha$; AS ONE Corporation, Japan), 
while heated water was supplied using a heating thermostat unit 
(HS-1, Tokyo Rikagaku Kikai Co., Ltd., Japan), with both streams conveyed by pumps.
The flow rate was adjusted using valves and Coriolis-type flowmeters (FD-SS2A; KEYENCE Corporation, Japan). 
Rod-type thermocouples were installed at the inlets and outlets on both the cold and hot sides via connectors, 
and steady-state water temperatures were measured after sufficient time had passed. 
In addition, the surface of the heat exchanger was covered with a thermal insulation material.
A differential pressure gauge (PZ-77; Tsukasa Sokken Co., Ltd., Japan) was used to measure the pressure drop through the heat exchanger.

\section{Results}

\subsection{Derivation of effective physical properties}
\begin{figure}[H]
\centering
\includegraphics[scale=0.9,clip]{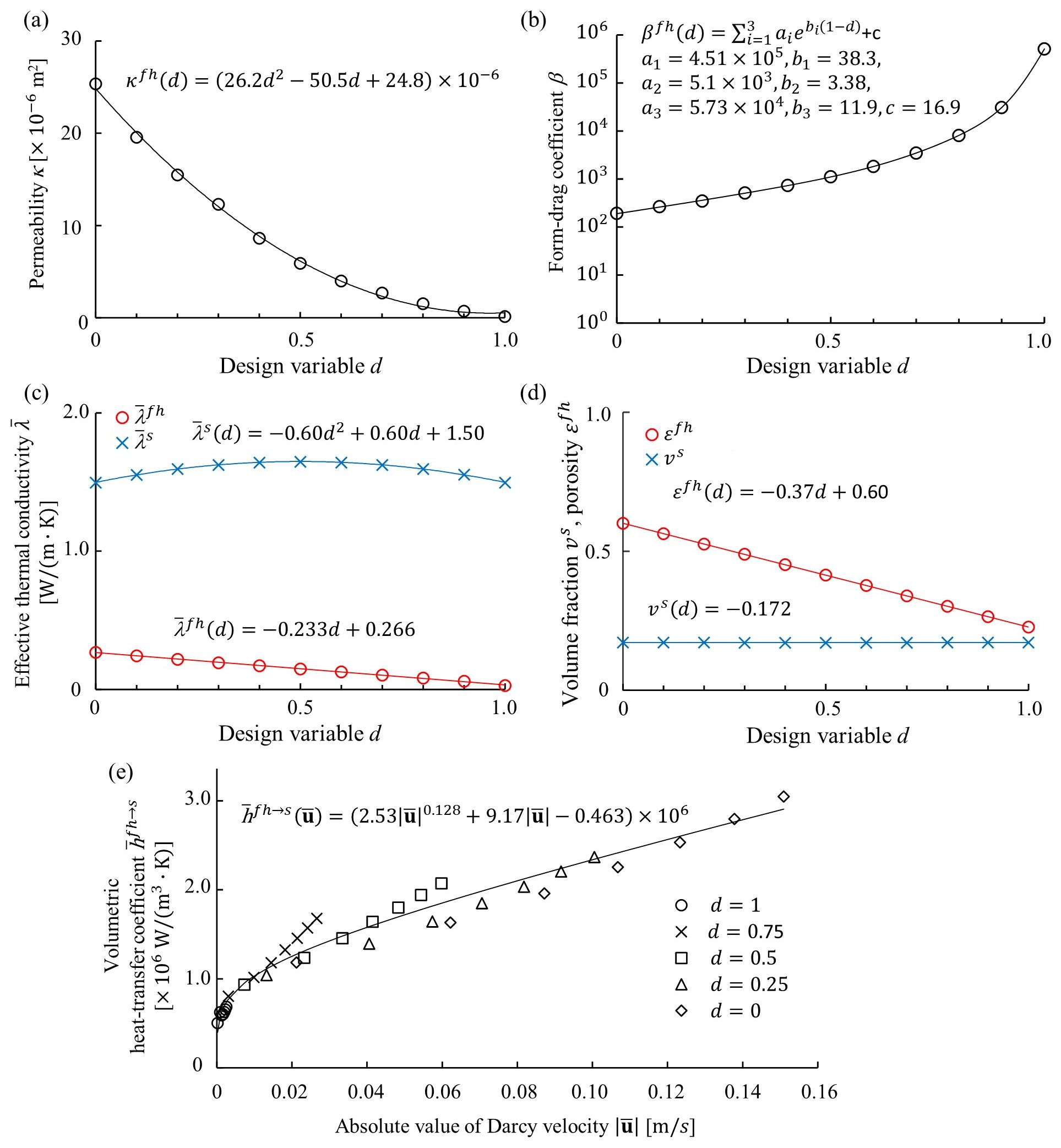}
\caption{Approximate representations of (a) permeability, (b) drag coefficient, (c) effective thermal conductivity, (d) porosity and solid volume fraction, and (e) the volumetric heat-transfer coefficient.}
\label{interp}
\end{figure}

Fig. \ref{interp} shows the approximate effective parameters evaluated using the unit cell depicted in Fig. \ref{rve}, as functions of the design variable $d$. In the definition of the unit cell geometry used in this study, the shape of the hot fluid region corresponding to the design variable $d$ is identical to that of the cold fluid region corresponding to $1-d$. In other words, the effective parameters of the hot fluid region for design variable $d$ are equal to those of the cold fluid region for design variable $1-d$. Therefore, only the effective material properties of the hot fluid region are presented here.
The pressure gradient $dP/dl$ applied to the fluid region was set to a maximum of 5 $\mathrm{kPa/m}$, 
assuming a pressure drop of 500 Pa over a channel length of approximately 0.1 m. The permeability $\kappa$ and drag coefficient $\beta$ are shown in Figs. \ref{interp} (a) and (b), the effective thermal conductivity $\overline{\lambda}$ in Fig. \ref{interp} (c), the porosity $\varepsilon$ and the volume fraction $v$ in Fig. \ref{interp} (d), and the volumetric heat-transfer coefficient $\overline{h}$ in Fig. \ref{interp} (e).

Figs. \ref{interp}(a), (b), and (e) indicate that as the design variable $d$ increases, the porosity of the hot fluid region decreases and the flow resistance increases, which is a physically reasonable trend. In addition, the volume fraction of the solid region remains nearly constant regardless of $d$, implying that the overall solid volume changes only slightly throughout the optimization. Therefore, no additional explicit global volume constraint was imposed in the present optimization problem. However, the bounds on the design variable defined through $C_{\min}$ and $C_{\max}$ implicitly restrict the admissible channel volume fractions by preventing either channel from becoming excessively narrow.

Furthermore, the volumetric heat-transfer coefficient shown in Fig.~\ref{interp}(e) is, in principle, an effective property that depends on both the design variable $d$ and the Darcy velocity $\bar{\bold{u}}$. In the present study, however, it was approximated as a function of the Darcy velocity alone. This simplification was adopted based on our RVE results, which indicated that, within the pressure-gradient range below $dP/dl = 5\,\mathrm{kPa/m}$ used to derive the effective properties, the dependence on $d$ was relatively small. In addition, explicitly incorporating the $d$-dependence into the optimization would increase the nonlinearity of the problem and make convergence more difficult. Therefore, to keep the optimization tractable, the volumetric heat-transfer coefficient was treated as a function only of the Darcy velocity.

\subsection{Optimization results based on the macroscopic model analysis}

\begin{figure}[H]
\centering
\includegraphics[scale=1.0,clip]{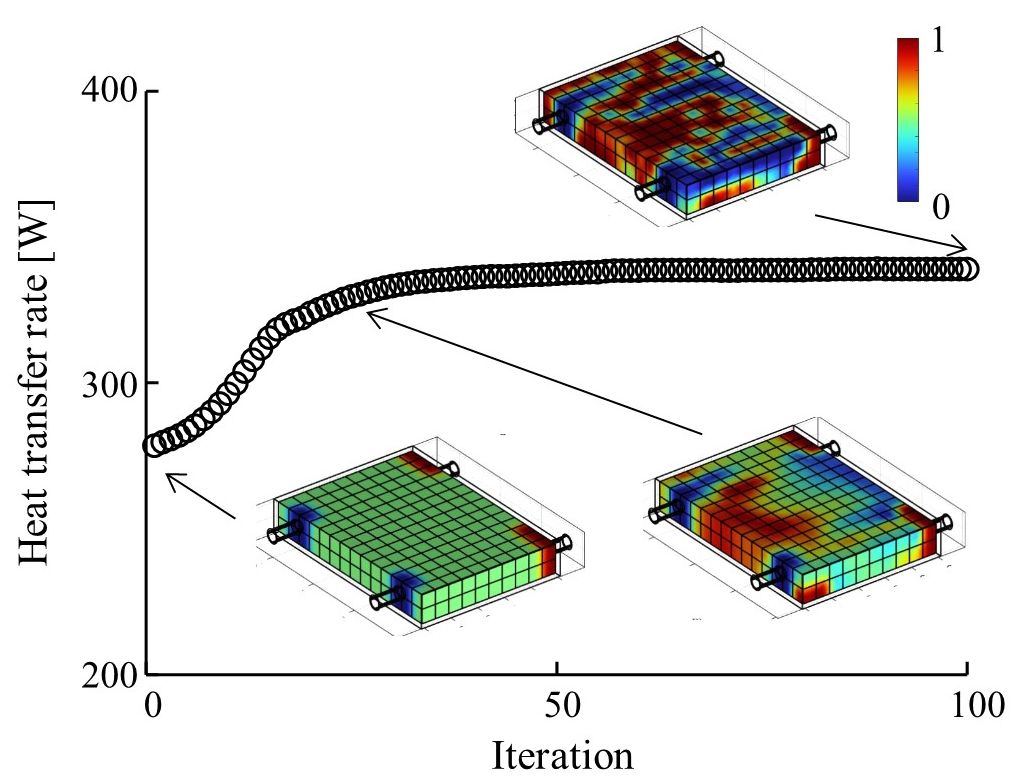}
\caption{Convergence history and corresponding intermediate design variable distributions.}
\label{history}
\end{figure}

Fig. \ref{history} illustrates the convergence history of the objective function. 
The optimization was terminated at 100 iterations after confirming that the change 
in the objective function is sufficiently small. 
The objective function converged stably and the heat transfer rate improved by approximately 20\%. 

\subsection{Re-analysis for the optimal result}

\begin{figure}[H]
\centering
\includegraphics[scale=1.0,clip]{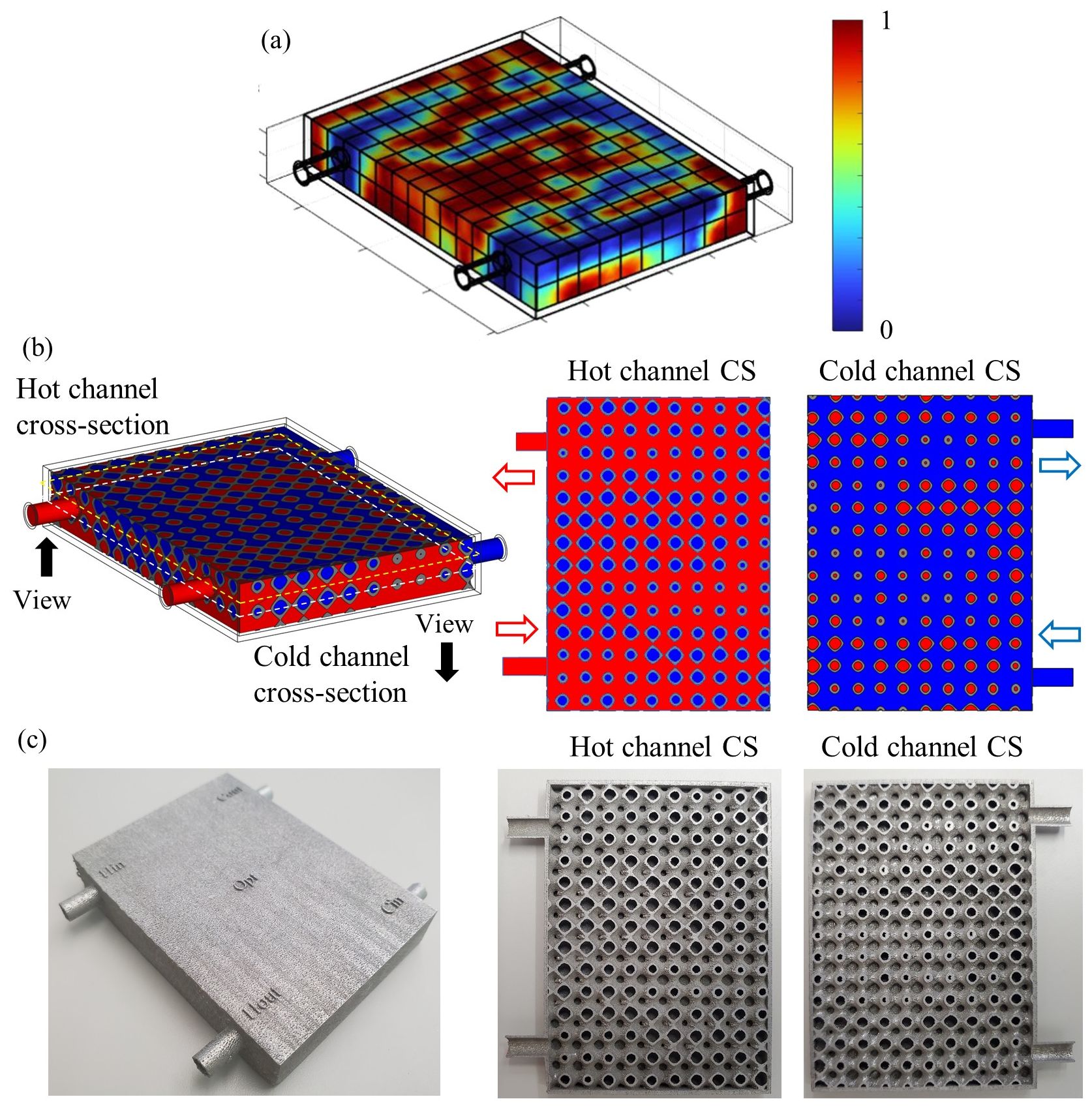}
\caption{Outlines of optimal results. (a) Optimal design variable distribution. (b) Detailed geometry generated from the optimal design variable distribution. (c) Fabricated test piece by metal LPBF.}
\label{optimalgeom}
\end{figure}

Fig. \ref{optimalgeom} (a) shows the distribution of optimal design variables,
and Fig. \ref{optimalgeom} (b) shows the detailed geometry generated from this distribution. 
The detailed geometry is illustrated in the cross-sectional views taken through the centers of the inlets and outlets of both the hot fluid and cold fluid regions. 
The distribution of the channel width clearly corresponds to the distribution of the design variables. Table \ref{surface_volume} compares the surface area of the fluid domain and the volume of the solid domain between the uniform lattice and the optimal lattice. Since the present study optimizes only the wall position, neither quantity changes significantly.
Fig. \ref{optimalgeom} (c) shows photographs of the fabricated heat exchanger, including its exterior and cross-sectional views captured through the hot- and cold-fluid regions. The fabricated geometry generally matches the original three-dimensional model data, and no cross-leakage between the channels was observed. To further examine the dimensional consistency at the product level, Table~\ref{masscomparison} compares the theoretical and measured overall masses of the fabricated specimens. The close agreement between these values supports the dimensional consistency of the fabricated heat exchangers.

\begin{table}[H]
\centering
\caption{Surface area of the fluid domain and volume of the solid domain of the TPMS lattice heat exchanger.}
\label{surface_volume}
\begin{tabular}{cccc}
\hline
 & \multicolumn{2}{c}{Surface area of fluid domain [mm$^2$]} & Volume [mm$^3$] \\
\cline{2-3}
 & Hot fluid & Cold fluid &  \\
\hline
Uniform lattice& 19312.5 & 18834.4 & 16247.0 \\
Optimal lattice& 18703.1 & 19128.8 & 16306.1 \\
\hline
\end{tabular}
\end{table}

\begin{table}[H]
\centering
\caption{Comparison of theoretical and measured specimen masses (316L density: 8.00 g/cm$^3$; assumed relative density: 99.0\%).}
\label{masscomparison}
\scriptsize
\begin{tabular}{lcccc}
\hline
 & \shortstack{Model volume\\{[cm$^3$]}} & \shortstack{Theoretical specimen\\ mass [g]} & \shortstack{Actual specimen\\ mass [g]}&\shortstack{Deviation of actual\\ specimen mass\\ from theoretical mass [\%]}\\
\hline
Uniform lattice   & 16.25 & 128.67 & 127.80 & -0.68 \\
Optimal lattice & 16.31 & 129.14 & 130.94 & 1.39 \\
\hline
\end{tabular}
\end{table}

\begin{figure}[H]
\centering
\includegraphics[scale=0.9,clip]{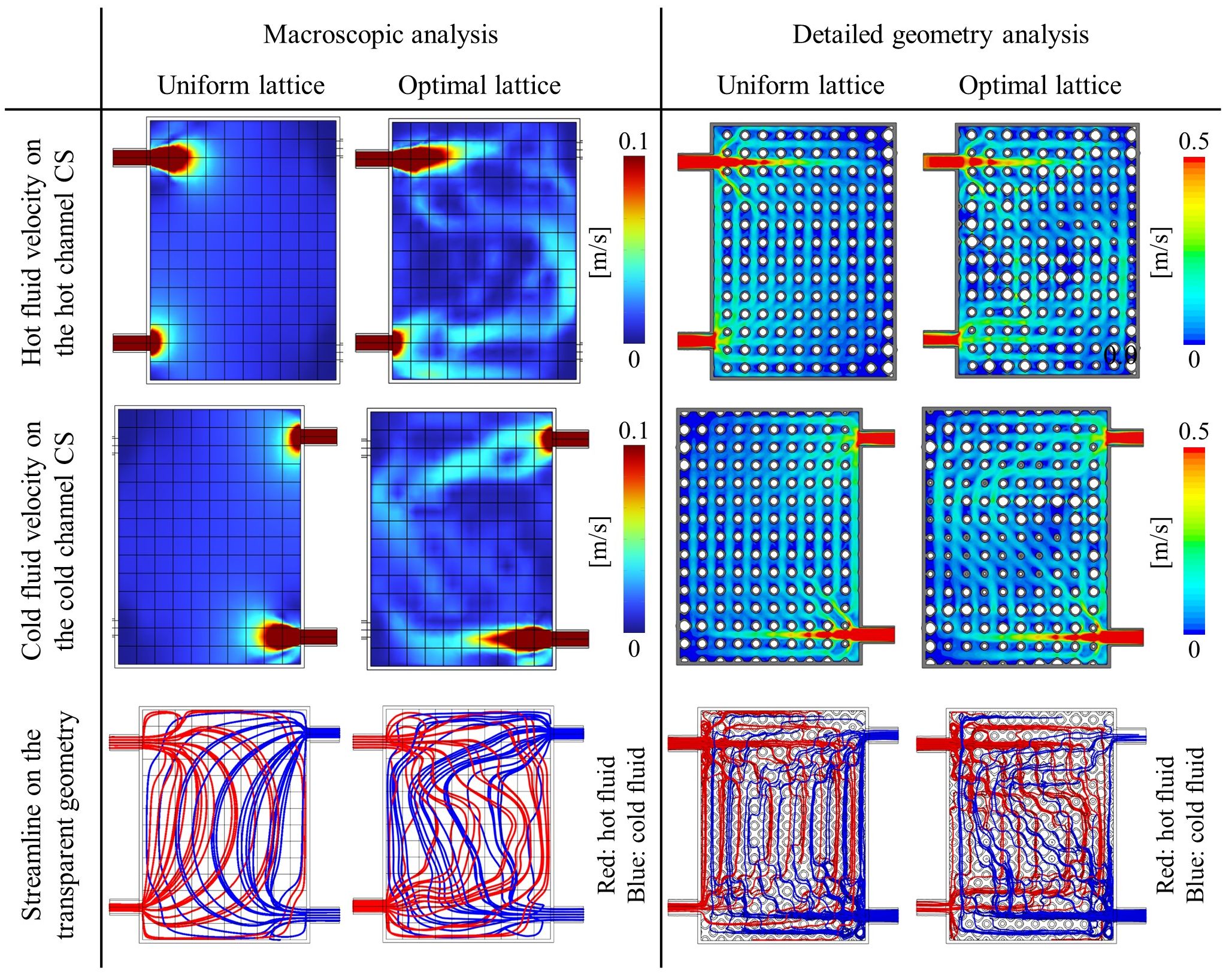}
\caption{Comparison of the hydrodynamic characteristics of the uniform and optimal lattices obtained from the macroscopic and the detailed-geometry analysis. CS is an abbreviation for Cross Section.}
\label{hydroanalysis}
\end{figure}

\begin{figure}[H]
\centering
\includegraphics[scale=0.9,clip]{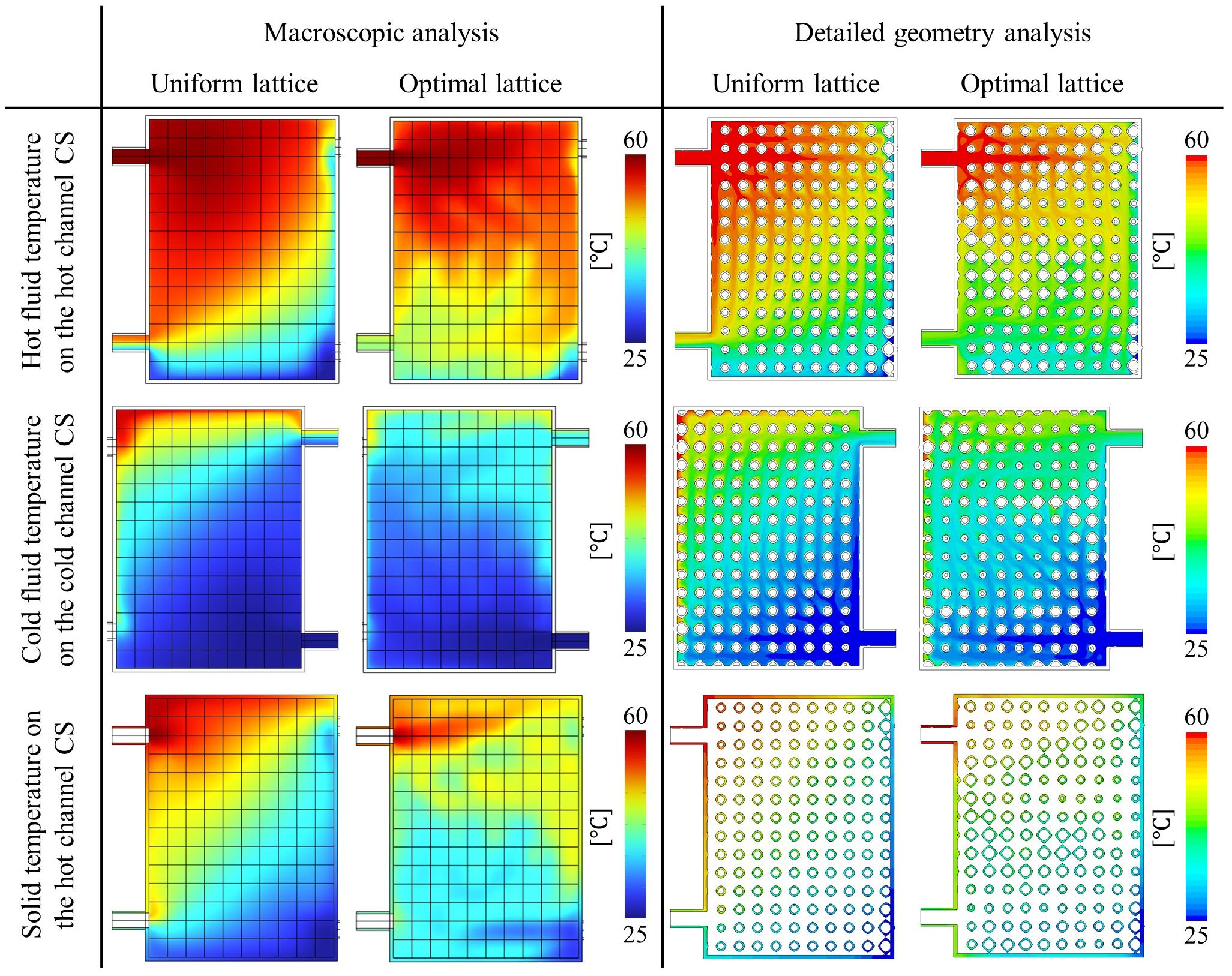}
\caption{Comparison of the thermal characteristics of the uniform and optimal lattices obtained from the macroscopic and the detailed-geometry analysis}
\label{thermalanalysis}
\end{figure}

Fig. \ref{hydroanalysis} compares the hydrodynamic characteristics of the uniform $(d=0)$ and optimal lattices obtained from both the macroscopic model and the detailed-geometry analysis, and similarly, Fig. \ref{thermalanalysis} compares their thermal characteristics. The uniform lattice was adopted as the benchmark because it is the configuration most widely studied in TPMS-based heat exchangers. In both comparisons, the results obtained using the macroscopic model 
and those obtained from the detailed geometry exhibit the same overall trends.

\begin{figure}[H]
\centering
\includegraphics[scale=0.9,clip]{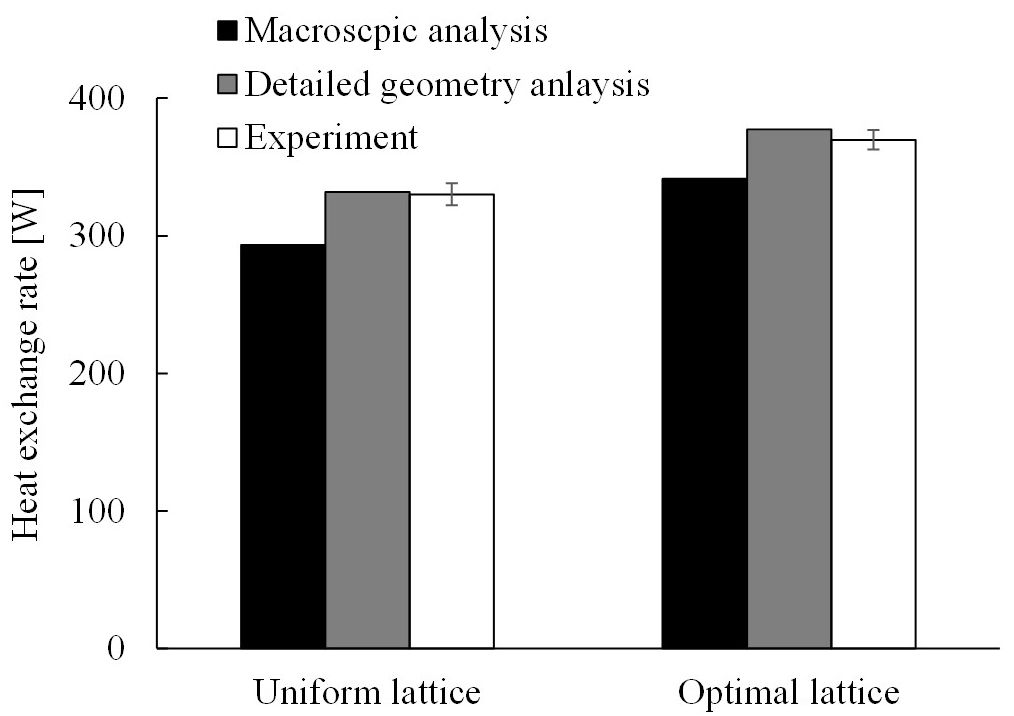}
\caption{Comparison of the heat transfer rate of the uniform and optimal lattices obtained from the macroscopic and the detailed-geometry analysis, and experiment.}
\label{heattransf}
\end{figure}

The flow rates, inlet-outlet temperature differences, and heat exchange rates are summarized in Table~\ref{heatexchangecomparison}, where the experimental results are presented as the mean \(\pm\) standard deviation based on repeated measurements \((N=3)\), together with the corresponding results from the macroscopic and detailed-geometry simulations for direct comparison. Fig.~\ref{heattransf} shows the heat transfer rates, which were evaluated in both the simulations and the experiments from the difference between the inlet and outlet enthalpy flow rates estimated from the mass flow rate and fluid temperature, and then averaged between the hot and cold sides. Although the detailed-geometry model yielded slightly higher heat transfer rates, the performance improvement from the uniform lattice to the optimal lattice was consistently reproduced. In the experiments, the inlet and outlet temperatures were measured using thermocouples, and the flow rates were obtained from the flow meter. The experimental results showed good agreement with the simulations, confirming that the optimization led to an increase in the heat transfer rate.

\begin{figure}[H]
\centering
\includegraphics[scale=0.8,clip]{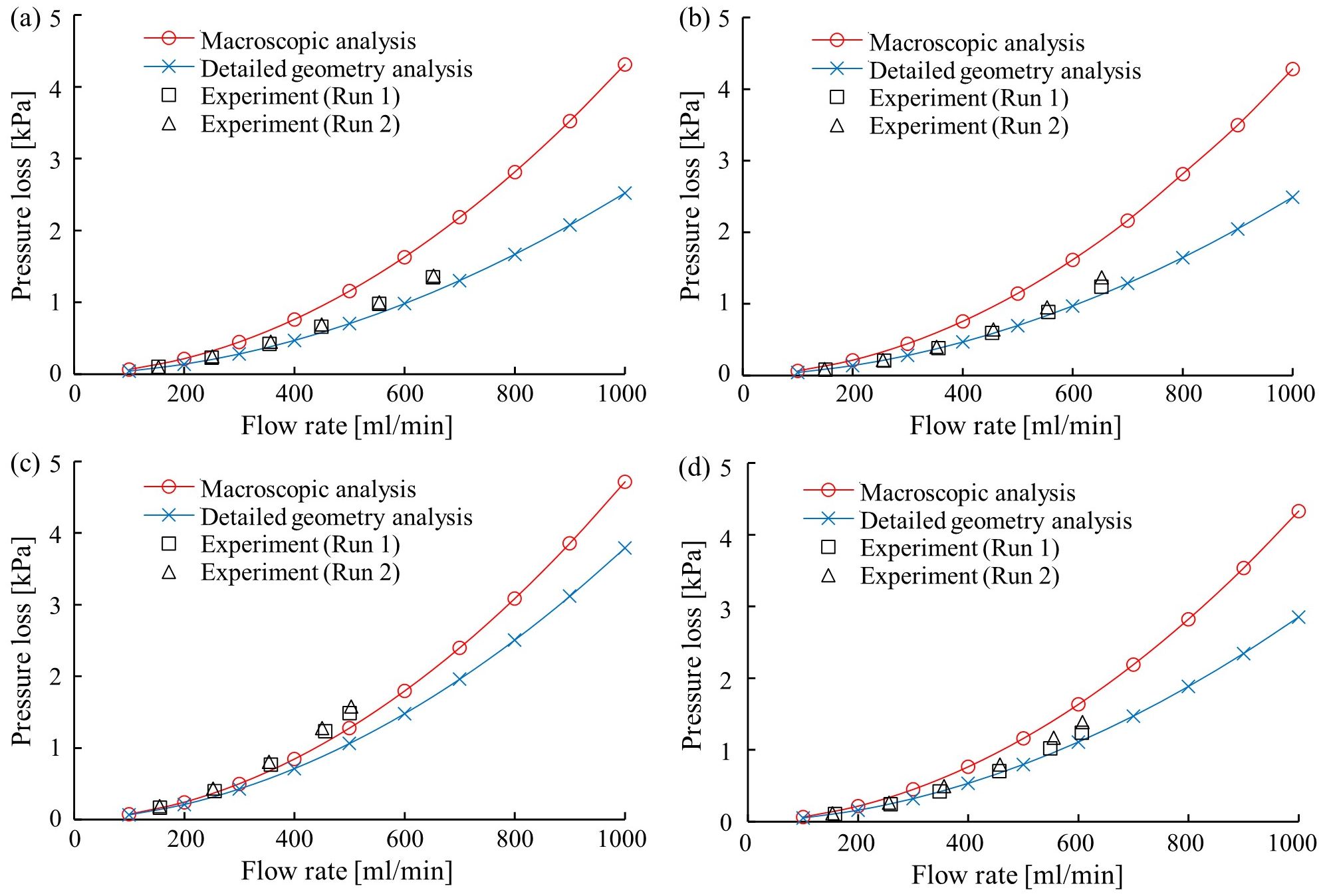}
\caption{Comparison of the pressure loss of the uniform and optimal lattices obtained from the macroscopic and the detailed-geometry analysis, and experiment. (a) Hot-fluid channel of the uniform lattice. (b) Cold-fluid channel of the uniform lattice. (c) Hot-fluid channel of the optimal lattice. (d) Cold-fluid channel of the optimal lattice.}
\label{pressure}
\end{figure}

Furthermore, Fig. \ref{pressure} presents a comparison of the pressure losses of the uniform and optimal lattices obtained from both the macroscopic model and the detailed-geometry analysis.
To enable comparison with the experimental results, the pressure losses were computed over a wide range of flow rates.
The pressure losses for the uniform and optimal lattices exhibit no significant differences, which is consistent with the small difference in the flow rates shown in Table \ref{heatexchangecomparison}.
However, the macroscopic model generally predicted higher pressure losses than the detailed model.

The measured differential pressures are presented in Fig. \ref{pressure}. Each experiment was repeated twice $(N=2)$. Because the maximum measurement range of the differential pressure gauge was 2 kPa, fewer experimental data points were available compared with the simulations.
In general, the experimental values were between those predicted by the macroscopic and detailed geometry models, except for the hot-fluid channel of the optimal lattice, where the experimental value was the highest.

\begin{table}[H]
\centering
\caption{Comparison of the flow rates, inlet--outlet temperature differences, and heat exchange rates of the uniform and optimal lattices obtained from the macroscopic analysis, detailed-geometry analysis, and experiments.}
\label{heatexchangecomparison}
\scriptsize
\setlength{\tabcolsep}{2pt}
\renewcommand{\arraystretch}{1.2}
\begin{tabular}{llccccccc}
\hline
& & \multicolumn{2}{c}{Flow rate [mL/min]}
& \multicolumn{2}{c}{\shortstack{Inlet--outlet temperature\\ difference [$^\circ$C]}}
& \multicolumn{3}{c}{Heat exchange rate [W]} \\
\cmidrule(lr){3-4} \cmidrule(lr){5-6} \cmidrule(lr){7-9}
& & Hot fluid & Cold fluid & Hot fluid & Cold fluid & Hot fluid & Cold fluid & Average \\
\hline
\multirow{2}{*}{\shortstack{Macroscopic\\simulation}}
& Uniform lattice & 319.2 & 320.9 & -13.2 & 13.0 & 294.9 & 292.0 & 293.5 \\
& Optimal lattice & 301.2 & 319.0 & -16.5 & 15.0 & 347.9 & 335.0 & 341.4 \\
\hline
\multirow{2}{*}{\shortstack{Detailed\\simulation}}
& Uniform lattice & 414.8 & 416.1 & -11.5 & 11.3 & 333.6 & 330.4 & 332.0 \\
& Optimal lattice & 328.5 & 386.7 & -16.3 & 14.1 & 374.4 & 380.6 & 377.5 \\
\hline
\multirow{2}{*}{Experiment}
& Uniform lattice & $409.3 \pm 5.5$ & $408.7 \pm 1.9$ & $-11.6 \pm 0.1$ & $11.5 \pm 0.6$ & $332.2 \pm 1.7$ & $328.3 \pm 17.2$ & $330.1 \pm 8.0$ \\
& Optimal lattice & $329.9 \pm 1.4$ & $396.1 \pm 0.5$ & $-16.1 \pm 0.6$ & $13.3 \pm 0.1$ & $371.8 \pm 12.7$ & $367.7 \pm 1.6$ & $369.8 \pm 7.1$ \\
\hline
\end{tabular}
\end{table}

\subsection{Computational cost comparison}

The computational times for one macro-scale analysis and one detailed analysis of the initial and optimal solutions are summarized in Table \ref{cptime}. The computational environment and solver configurations are also listed therein. In analyses employing detailed geometries, although the FVM is generally regarded as being faster than the FEM, the computational time still amounted to nearly three hours per run. In contrast, the macro-scale analysis achieved a 92.8\% reduction in computational time. Because different numerical methods were employed in the macro-scale and detailed-geometry analyses, a rigorous quantitative comparison of the computational cost reduction is not possible. Nevertheless, it is clear that the computational cost was reduced to a level at which iterative optimization procedures become practically feasible.

\begin{table}[t]
\centering
\caption{Computational settings and times for macroscopic and detailed-geometry analyses}
\label{cptime}
\footnotesize
\begin{tabular}{lp{5cm}p{5cm}}
\hline
 & Macroscopic analysis & Detailed-geometry analysis \\
\hline
CPU & \multicolumn{2}{c}{Intel(R) Xeon(R) Gold 5222 CPU @ 3.80\,GHz} \\
Memory & \multicolumn{2}{c}{64\,GB} \\
Numerical method & FEM & FVM \\
Meshing strategy & Tetrahedral mesh with three prism layers applied at the wall boundaries (maximum cell size: 3 mm; prism-layer thickness: 1 mm). &Polyhedral mesh with two prism layers applied at the wall boundaries (maximum cell size: 1 mm; prism-layer thickness: 0.3 mm).\\ 
Number of cells & 72904 & 1365178 (uniform lattice) and 1363066 (optimal lattice)\\
Spatial discretization & First-order Lagrange element & Second-order finite volume discretization \\
Coupling strategy & Fully coupled (monolithic) & Coupled pressure--velocity solver\\
Linear solver & PARDISO (direct) & AMG (V-cycle) \\
Computational time & 686.2\,s & 9484.4\,s \\
\hline
\end{tabular}
\end{table}

\section{Discussion}
First, with respect to the optimal solution, the design variable distribution and streamlines shown in Fig. \ref{hydroanalysis} indicate that, while the flow in the uniform lattice develops predominantly in the vertical direction for both the hot and cold fluids, the optimal solution induces the flow in a diagonal direction. Therefore, it can be inferred that optimization increased the interaction length between the two fluids. This approach of extending the distance over which the flow paths intersect was also implemented by Oh et al. through the introduction of internal walls \cite{oh2023}, and it can be considered a rational method for enhancing the heat exchange performance.

Here, a quantitative evaluation of the effective flow path length is performed. 
Assuming that the pressure loss $\Delta P$ can be expressed as the product of the effective flow path length $L_{\mathrm{eff}}$ and the flow-direction-averaged pressure gradient $\overline{-\nabla P \cdot \mathbf{e}}$, where $\mathbf{e}=\frac{\mathbf{u}}{|\mathbf{u}|}$ denotes the unit vector in the local flow direction, the effective flow path length is defined as
\begin{equation}
L_{\mathrm{eff}}=\frac{\Delta P}{\overline{-\nabla P \cdot \mathbf{e}}},
\end{equation}
For the complex flow field inside the heat exchanger, the averaged quantity $\overline{-\nabla P \cdot \mathbf{e}}$ is evaluated as the mass-flux-weighted volume average \cite{whitaker1999} given by
\begin{equation}
\overline{-\nabla P \cdot \mathbf{e}}=
\frac{\displaystyle\int_{\Omega} (-\nabla P \cdot \mathbf{e})\, \rho |\mathbf{u}| \, dV}
{\displaystyle\int_{\Omega} \rho |\mathbf{u}| \, dV}.
\end{equation}
The resulting effective flow path lengths obtained for the optimal lattice and the uniform lattice are summarized in Table \ref{leff}. The effective flow path lengths of the optimal lattice are longer than those of the uniform lattice.

\begin{table}[H]
\centering
\caption{Comparison of the effective flow path length between the uniform and optimal lattice structures.}
\label{leff}
\begin{tabular}{c c c}
\hline
Effective flow path length [m] & Hot fluid & Cold fluid \\ \hline \hline
Uniform lattice  & 0.134 & 0.141 \\
Optimal lattice & 0.169 & 0.166 \\
\hline
\end{tabular}
\end{table}

In addition, the velocity distribution exhibited a clearer variation in the optimal lattice than in the uniform lattice. Oh et al. argued that a uniform velocity distribution is optimal for heat-exchange efficiency in their optimization of graded lattice geometries \cite{oh2025}. The study by Oh et al. considered a cuboid heat exchanger in which a U-shaped counterflow was introduced diagonally from the same face. In that configuration, a clear dead zone of the flow was generated, and therefore directing the flow into that region naturally became an optimal design strategy. In contrast, the heat exchanger investigated in the present study introduces a U-shaped counterflow in a planar manner, where the formation of flow dead zones is inherently less likely. Thus, the presence or absence of such dead zones may be one of the factors leading to the different conclusion that a non-uniform flow distribution becomes optimal in this study.

These advantages in the flow behavior are reflected in the temperature distributions of each fluid, as shown in Fig. \ref{thermalanalysis}. For example, in the uniform lattice, the flow near the left wall became stronger from the perspective of the hot-fluid channel because this region is close to both the inlet and outlet of the hot channel. In contrast, on the cold-fluid side, this region is farther from the inlet and outlet; therefore, the influence of the hot-fluid flow becomes more pronounced, and the temperature decreases less readily. Near the right wall, the opposite behavior occurred. As a result, a temperature imbalance was generated, that is, higher temperatures near the left wall and lower temperatures near the right wall. However, in the optimal lattice, the induced diagonal crossing of the flow reduced the temperature difference between the left and right sides of the heat exchanger. Such a flat and unbiased temperature distribution is typically observed in straight heat exchangers; however, it is difficult to achieve this in more complex configurations such as U-shaped flows. Therefore, this result can be regarded as a clear benefit of the numerical optimization.

These characteristics were generally consistent between the macroscopic and detailed-geometry analyses in terms of the overall flow and temperature distributions, although some differences appeared in the detailed flow structures. In the hot-fluid channel, both analyses show only a small near-wall flow that directly connects the inlet and outlet. In contrast, in the cold-fluid channel such a flow does not appear in the macroscopic analysis, whereas the detailed-geometry analysis exhibits a preferential near-wall flow path connecting the inlet and outlet. Consequently, the flow toward the opposite wall becomes weaker, and the temperature distribution also differs: the low-temperature region spreads toward the far side in the macroscopic analysis, whereas it tends to spread toward the near side in the detailed-geometry analysis. Two possible reasons may explain this discrepancy: one is that the accuracy of the macroscopic flow model may deteriorate in the near-wall region, and the other is that the conversion from the design-variable distribution to the detailed geometry may have been inappropriate near the wall. Despite these discrepancies, and the quantitative differences in the heat transfer rate (Fig.~\ref{heattransf}) and pressure loss (Fig.~\ref{pressure}), the macroscopic model can still be regarded as sufficiently reliable for qualitative flow-priority optimization.

However, the macroscopic model clearly predicted higher pressure losses than the detailed model; a similar tendency was observed in our previous studies. In Ref. \cite{yanagihara2025}, the authors attributed the overestimation of pressure loss in the macroscopic model to the fact that the effective material properties were derived under the assumption of perfect lattice periodicity, while the optimal lattice becomes non-periodic, thereby invalidating the applicability of those effective properties. To further investigate this issue, a comparison of the analysis accuracy of the RVE for uniform and non-uniform structures was conducted, and the results are presented in the Appendix. The results indicate that, for the RVEs related to flow and heat conduction, the accuracy becomes higher when the structure is uniform. In contrast, the RVE for volumetric heat transfer inherently tends to underestimate the response, and structural non-uniformity partially alleviates this tendency. It should also be noted that, in the present study, the derivation of the volumetric heat transfer coefficient involves a simplifying assumption that the wall temperature is constant within each cell, and that the heat exchange within the cell is proportional to the difference between the mean fluid temperature and the wall temperature. Since this assumption neglects the temperature gradient within the solid wall, it does not correspond to a full treatment of conjugate heat transfer (CHT) at the cell scale. Although this simplification is currently necessary for constructing the macroscopic model, it may also constitute one of the fundamental sources of modeling error.

Furthermore, as shown by the pressure-loss results in Fig. \ref{pressure}, the pressure loss is overestimated even for the uniform lattice. This suggests that the discrepancy cannot be fully explained solely by the error evaluation for straight-flow cases such as those examined in the Appendix. Therefore, additional factors, such as the complex flow paths inherent in the heat-exchanger geometry and the deterioration of the macroscopic model accuracy near the wall, are also likely to contribute to the observed discrepancy. Further investigation is required to clarify these effects and to improve the predictive accuracy of the macroscopic model.

Figs. \ref{heattransf} and \ref{pressure} show that the experimental results agreed well with those of the detailed-geometry analysis, thereby confirming that the optimal solution is practically attainable. 
Since the heat transfer rates obtained from the detailed-geometry analysis and the experiments are similar, the influence of the specimen surface roughness on the heat transfer performance is considered to be small. However, in Fig. \ref{pressure}, only the hot-fluid channel of the optimal lattice exhibits a significantly larger pressure loss in the experiment than in the simulations. To investigate the reason for this, we first note that in the optimal lattice, the optimization of the flow priority can result in the throat of one of the fluid channels becoming extremely narrow, with a minimum diameter as small as $\phi 0.5\,\mathrm{mm}$. Because the shape resolution of LPBF is fixed by the equipment and processing conditions, the smaller the fabricated geometry, the greater the risk of geometric inaccuracies \cite{takezawa2017am}. Although the fabricated specimen shown in Fig. \ref{optimalgeom} (c) generally reproduces the 3D model, geometric imperfections can be observed when focusing on the throat region.

\begin{figure}[H]
\centering
\includegraphics[scale=0.9,clip]{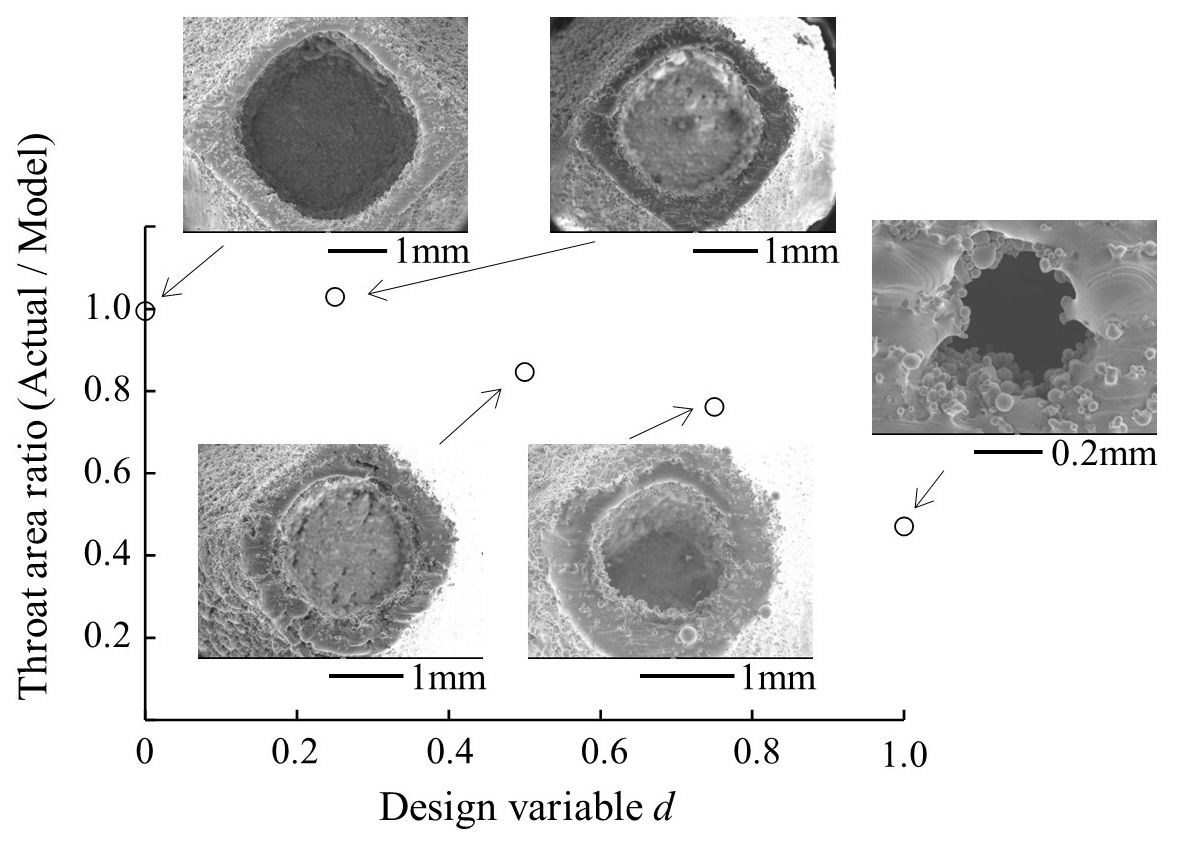}
\caption{Ratio of the TPMS lattice throat area measured from SEM images to the corresponding model area.}
\label{sem}
\end{figure}

Fig. \ref{sem} presents the results obtained by fabricating unit cells corresponding to five design variables ranging from $d=0$ to $1$ under the same processing conditions and build orientation as the test specimen. The throat region of the hot-fluid channel was observed using a scanning electron microscope (SEM), and the throat area was evaluated from the SEM images using a polygonal approximation and compared with that of the corresponding 3D model. Up to $d=0.4$, the measured throat area agrees well with the theoretical value. However, as $d$ increases beyond this value, the ratio gradually decreases, and the actual throat area eventually becomes less than half of that predicted by the model. As a result, small geometric inaccuracies significantly reduce the effective throat area. This interpretation is also consistent with the mass comparison summarized in Table~\ref{masscomparison}. Whereas the uniform lattice specimen exhibited a slightly smaller mass than the theoretical value, the optimal lattice specimen was 1.39\% heavier than the corresponding theoretical mass calculated from the model volume. This result suggests that excess material was sintered within the fabricated optimal lattice. Such imperfections are likely to have a more severe impact on the hot-fluid channel, leading to a significantly larger pressure loss. This issue can be mitigated either by increasing the minimum throat area through narrowing the range between $C_{\max}$ and $C_{\min}$ in Eq. \eqref{eq17}, or by increasing the size of the reference unit cell, which was set to 5~mm in the present study.

\begin{figure}[H]
\centering
\includegraphics[scale=0.8,clip]{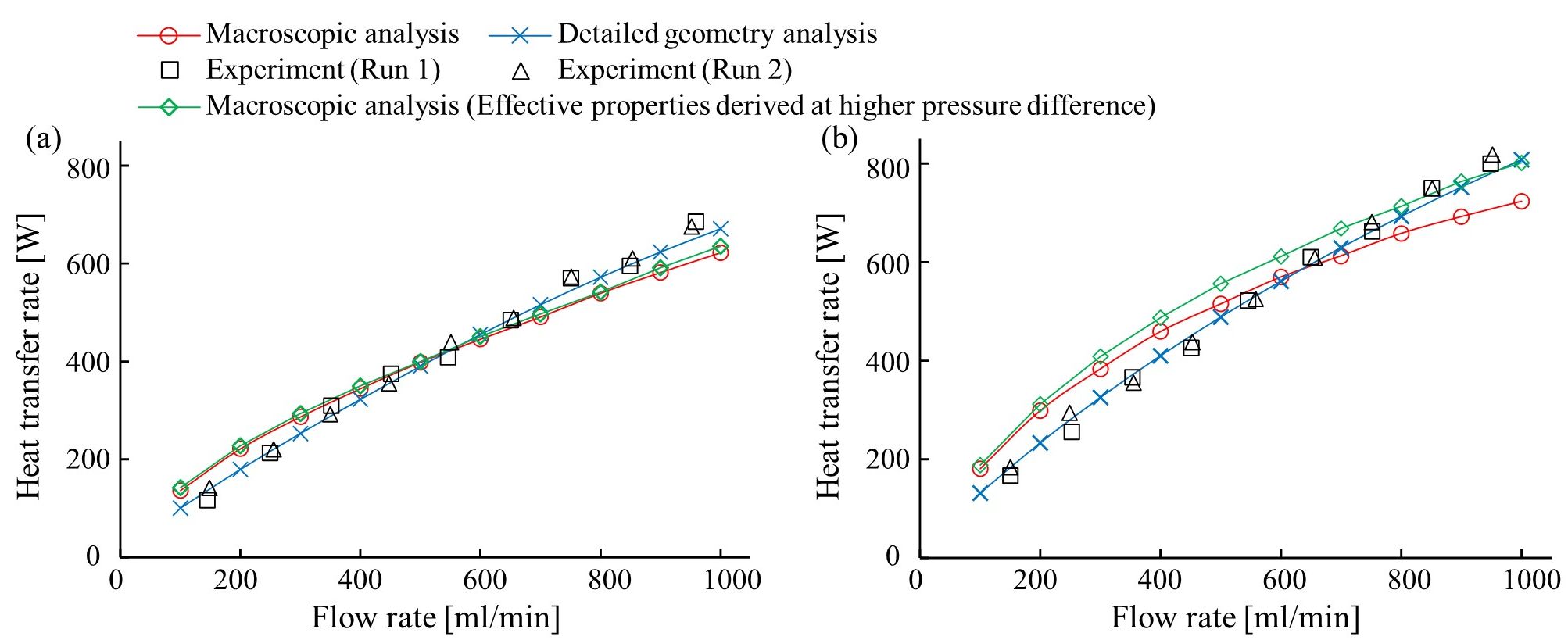}
\caption{Comparison of heat exchange rate under equal flow-rate conditions for the hot and cold fluids. (a) Uniform lattice. (b) Optimal lattice}
\label{comparison}
\end{figure}

Finally, because the optimization in this study was performed under the condition of a fixed pressure difference applied between the inlet and outlet, the resulting optimal solution yielded different flow rates on the hot and cold sides. For a simpler performance evaluation, both flow rates were equalized, and the system was re-evaluated; the results are shown in Fig. \ref{comparison}. Macroscopic simulations, detailed-geometry simulations, and experiments were conducted under this equal-flow condition. Each experiment was repeated twice $(N=2)$. In the macroscopic analysis performed here, the volumetric heat transfer coefficient $\bar{h}$ was not treated as a function solely of the Darcy velocity $\bar{\bold{u}}$, as shown in Fig. \ref{interp}(e) and used in the optimization. Instead, it was modeled as a two-variable distribution, where the dependence on the design variable $d$ was also taken into account. The heat transfer rate was derived using the same method as that used to obtain the data shown in Fig. \ref{heattransf}.

Based on a comparison of the heat exchange performance at the same flow rate, the optimal lattice exhibited an average improvement of 24.2\% in the detailed-geometry simulations and 23.3\% in the experimental results compared to the uniform lattice. This indicates that the structure obtained in this study is clearly superior to the uniform lattice, even when the flow rates differ slightly from those used to obtain the optimal solution and are adjusted to be equal on the hot and cold sides.

\begin{figure}[H]
\centering
\includegraphics[scale=0.8,clip]{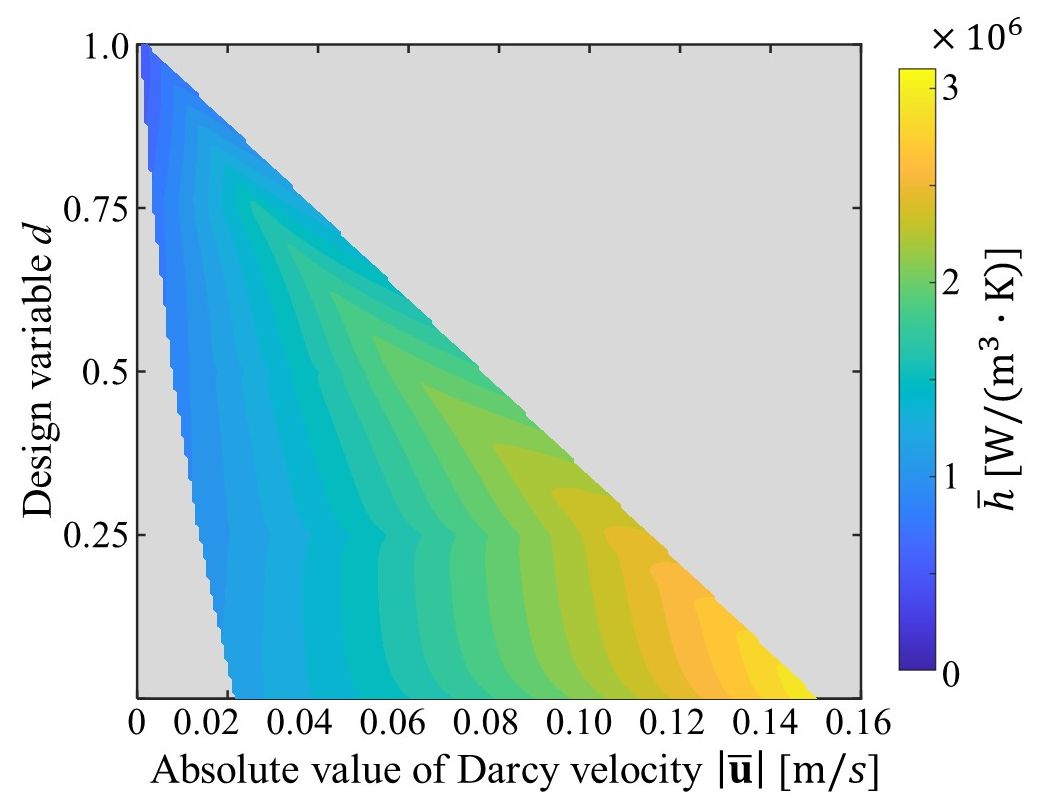}
\caption{Volumetric heat transfer coefficient distribution as a function of the design variable and Darcy velocity used in the re-analysis. The gray region indicates parameter combinations for which no simulation data are available. When parameter values outside the available data range are required in the re-analysis, the nearest available value is assigned as a numerical treatment.}
\label{hvcontour}
\end{figure}

It can also be observed that the discrepancy between the macroscopic and detailed geometry analyses increases as the flow rate increases. A similar trend was observed in the pressure loss results shown in Fig. \ref{pressure}, suggesting that the error originates from the flow analysis. 

The reason for this flow rate dependent error in the macroscopic flow model is that the effective parameters used in the macroscopic analysis were derived by applying a pressure difference of 500 Pa over a channel length of 0.1 m, as described in Section 3.1. This pressure difference corresponds to a flow rate of approximately 300--400 mL/min. In other words, the model yielded good accuracy within the flow-rate range. However, as the flow rate increases, the effective parameters deteriorate, leading to reduced accuracy.

To examine this point, the pressure difference used to derive the effective parameters was doubled (assuming a pressure difference of 1 kPa over a channel length of 0.1 m), and the results of the macroscopic analysis using these newly derived effective properties were added to Fig. \ref{comparison}. The results show that almost no difference was observed for the uniform lattice, whereas an improvement in accuracy in the high flow-rate region was observed only for the optimal lattice. This indicates that the discrepancy cannot be explained solely by the pressure-difference condition used in deriving the effective parameters. Rather, the results suggest that this issue is also closely related to the accuracy of the RVE when applied to spatially non-uniform lattice structures, as discussed earlier. Therefore, a more comprehensive investigation that simultaneously considers the flow-rate dependence of the effective parameters and the spatial non-periodicity of the lattice structure will be required.

Furthermore, if the present framework is to be used to obtain an optimal design that exhibits good performance over a wide range of flow rates, the computational cost would increase in proportion to the number of models employed. One possible approach is to prepare several models in which the effective parameters are derived for different pressure ranges, perform analyses for all these models, and treat the obtained results within a multi-objective optimization framework.

\section{Conclusion}

In this paper, we propose a macroscopic analysis model for a two-fluid heat exchanger 
by incorporating a TPMS Primitive lattice. 
Macroscopic flow analysis was conducted based on the Darcy--Forchheimer theory. 
Assuming that heat transfer occurs only between the fluid and the TPMS walls, 
we formulated a macroscopic heat transfer model by introducing a volumetric heat-transfer coefficient representing the macroscopic heat transfer rate per unit volume.

To optimize the priority between the hot and cold flows within the heat exchanger, namely, the channel widths, we employed the isosurface threshold of the primitive lattice as the design variable and developed a lattice-distribution optimization algorithm based on the aforementioned macroscopic analysis model. Optimization was then performed for a planar heat exchanger in which both the hot and cold fluids formed U-shaped flow paths. A stable optimal solution was obtained, and its validity was examined through detailed-geometry analysis and experiments conducted using a metal LPBF. The optimal solution derived from the macroscopic model also demonstrated a clear performance advantage over a uniform lattice in the experimental results. A physical interpretation of this improvement is that the optimization effectively increases the interaction length between the hot and cold fluids, enabling more uniform temperature distributions in each fluid.
On the other hand, conventional design theories, such as thermal resistance balancing, have been well established for heat exchangers with simple geometries. To clearly elucidate the physical mechanisms underlying the optimization process in the present study, a comparison with these conventional design theories will be an important direction for further research.

However, several limitations of the proposed macroscopic analysis and optimization methods have been identified. The accuracy of the macroscopic analysis depends on the pressure-gradient conditions under which the effective parameters were derived; therefore, it becomes difficult to maintain accuracy when the pressure or flow rate varies over a wide range. Furthermore, when the optimization produces lattices with narrow throats, geometric imperfections can occur during fabrication, leading to significantly increased pressure loss in the experiments. In the derivation of the effective parameters in the present study, several assumptions were introduced, such as the periodicity of the unit-cell geometry and the assumption of a constant solid temperature when deriving the volumetric heat-transfer coefficient. These assumptions may influence the accuracy of the macroscopic analysis. Improving the predictive accuracy by accounting for geometric and state non-periodicity or non-uniformity in the derivation of the effective parameters therefore remains an important challenge.

Furthermore, the fact that the macroscopic analysis overestimated the pressure loss regardless of the structural non-periodicity suggests that the macroscopic model inherently contains errors when applied to complex flow paths. Since this discrepancy becomes larger as the flow velocity increases, resolving this issue will be particularly important when extending the present approach to heat exchangers operating at higher Reynolds numbers.

From a practical standpoint, very small and fine cells may cause fouling problems, which may motivate increasing the cell size to some extent. However, excessively large cell sizes may weaken the self-supporting capability of the TPMS lattice and potentially lead to new geometric inaccuracies. Determining an appropriate cell size therefore remains an important design issue.

Finally, the optimization of the flow priority is a general methodology that can be applied not only to primitive lattices but also to other TPMS lattices and is not limited to U-shaped counterflow channels. Demonstrating the versatility of the proposed method through its application to various heat exchangers with different flow-path configurations, such as Z-type heat exchangers, as well as to different TPMS lattices, will be an important step toward broader applicability.

\section*{Acknowledgments}
This study was partly funded by JSPS KAKENHI (Grant No. 23H01324) and the Paloma Environmental Technology Development Foundation.

\appendix
\section{Effect of spatial nonuniformity on the macroscopic parameters' accuracy}

The error observed in the macroscopic analysis in this study is attributed to the fact that the effective parameters derived using the RVE method are based on the assumption that the unit cells are arranged in a perfectly periodic manner. However, in the optimal solution obtained in this study, this periodicity is significantly disrupted.

\begin{figure}[H]
\centering
\includegraphics[scale=1.0,clip]{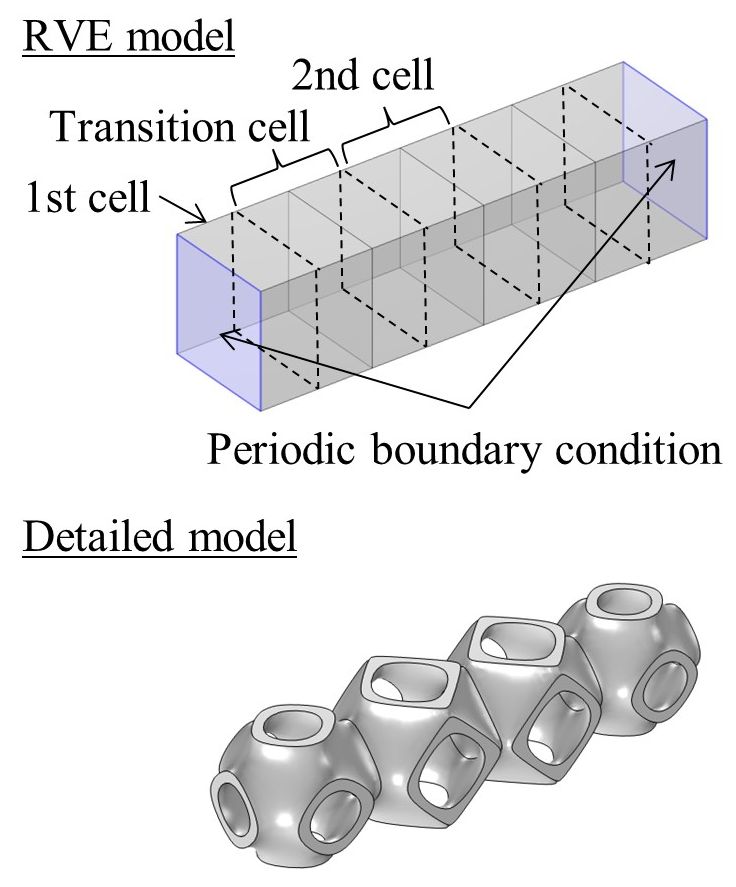}
\caption{Analysis model for evaluating the effect of spatial nonuniformity on the accuracy of the RVE. All lateral boundary conditions not explicitly indicated in the figure are periodic boundary conditions.}
\label{RVEcheckmodel}
\end{figure}

To investigate the influence of such geometric non-periodicity on the accuracy of the macroscopic model, a four-cell model shown in Fig.~\ref{RVEcheckmodel} is considered in this appendix. The model consists of the first cell, the second cell, and two transition cells located between them.

Let the design variables of the first and second cells be $d_1$ and $d_2$, respectively. The design variables of the transition cells are determined by linearly interpolating between $d_1$ and $d_2$. Using both the macroscopic model and the detailed model, the following quantities are evaluated: the Darcy velocity of the hot fluid under the boundary conditions shown in Fig.~\ref{rve}(a), the heat flux in the solid region and the hot fluid region under the boundary conditions shown in Fig.~\ref{rve}(b), and the volumetric heat generation under the boundary conditions shown in Fig.~\ref{rve}(c). The error of the macroscopic model with respect to the detailed model is then calculated for multiple combinations of $d_1$ and $d_2$.

\begin{figure}[H]
\centering
\includegraphics[scale=1.0,clip]{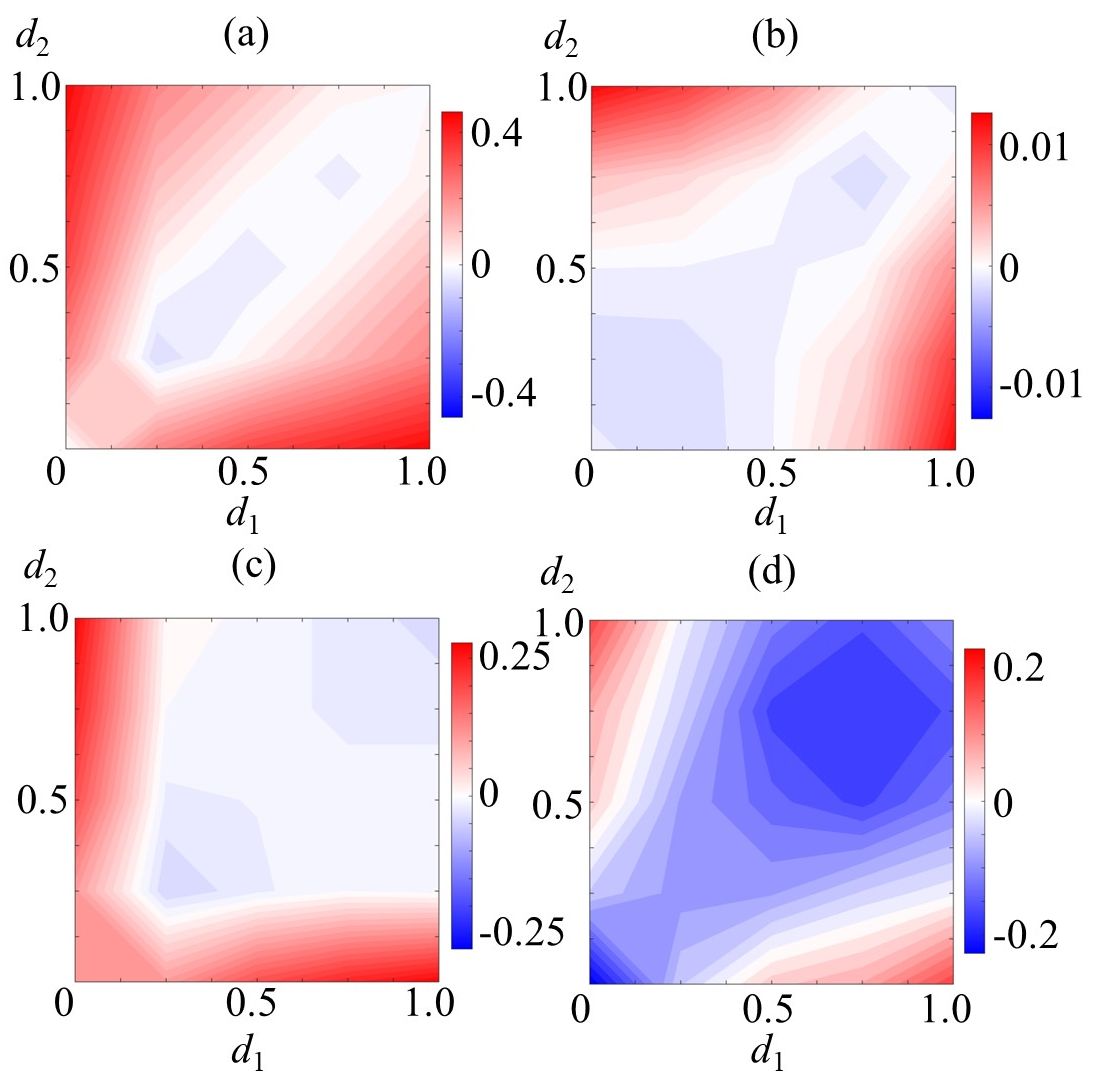}
\caption{Error of the RVE model relative to the detailed model: (a) Darcy velocity of the hot fluid, (b) heat flux in the solid domain, (c) heat flux in the hot fluid domain, and (d) volumetric heat generation.}
\label{RVEcheckresults}
\end{figure}

The results are shown in Fig.~\ref{RVEcheckresults}. For the Darcy velocity in Fig.~\ref{RVEcheckresults}(a) and the heat flux in Fig.~\ref{RVEcheckresults}(b) and (c), the accuracy is highest when $d_1 = d_2$, and the error increases as the difference between $d_1$ and $d_2$ becomes larger. This clearly indicates that the breakdown of periodicity directly leads to a deterioration of the macroscopic model accuracy.

In contrast, for the volumetric heat generation shown in Fig.~\ref{RVEcheckresults}(d), the macroscopic model underestimates the heat generation when $d_1 = d_2$, and the minimum error appears in a region where a moderate difference between $d_1$ and $d_2$ exists. This behavior can be interpreted as follows. Volumetric heat generation is closely related to the Darcy velocity and tends to be intrinsically underestimated for TPMS lattices. When the periodicity is disrupted, the overestimation of Darcy velocity induced by the non-periodic geometry partially compensates for this underestimation, resulting in the observed reduction in the error.

These results suggest that the influence of spatial non-periodicity on RVE accuracy depends on the physical quantity being evaluated. While periodicity is essential for accurately predicting flow and heat flux, its breakdown may partially compensate for the inherent underestimation in volumetric heat generation. Therefore, careful consideration of spatial non-uniformity is required when applying RVE-based homogenization to optimal lattice structures.

\end{document}